\pdfoutput=1
\documentclass[%
reprint,
superscriptaddress,
amsmath,amssymb,
aps,
floatfix,
twocolumn]{revtex4-1}
\usepackage{graphicx}
\usepackage{dcolumn}
\usepackage{bm}
\usepackage{amsmath}
\usepackage{amssymb}
\usepackage[colorlinks,
linkcolor=red,
anchorcolor=red,
citecolor=red
]{hyperref}

\usepackage{mathtools}
\DeclarePairedDelimiter\bra{\langle}{\rvert}
\DeclarePairedDelimiter\ket{\lvert}{\rangle}
\DeclarePairedDelimiterX\braket[2]{\langle}{\rangle}{#1 \delimsize\vert #2}
\begin{document}
	%
	
	\title{Transport study of the wormhole effect in three-dimensional topological insulators}
	
	\author{Ming Gong}
	\affiliation{International Center for Quantum Materials, School of Physics, Peking University, Beijing 100871, China}
	\author{Ming Lu}
	\affiliation{Beijing Academy of Quantum Information Sciences, West Bld.3,
		No.10 Xibeiwang East Rd., Haidian District, Beijing 100193, China}
	\affiliation{International Center for Quantum Materials, School of Physics, Peking University, Beijing 100871, China}
	\author{Haiwen Liu}
	\affiliation{Center for Advanced Quantum Studies, Department of Physics, Beijing Normal University, Beijing 100875, China}
	\author{Hua Jiang}\thanks{jianghuaphy@suda.edu.cn}
	\affiliation{School of Physical Science and Technology, Soochow University, Suzhou, 215006, China.}
	\affiliation{Institute for Advanced Study and School of Physical Science and Technology, Soochow University, Suzhou 215006, China.}
	\author{Qing-Feng Sun}
	\affiliation{International Center for Quantum Materials, School of Physics, Peking University, Beijing 100871, China}
	\affiliation{Collaborative Innovation Center of Quantum Matter, Beijing 100871, China}
	\affiliation{CAS Center for Excellence in Topological Quantum Computation,
		University of Chinese Academy of Sciences, Beijing 100190, China}
	\author{X. C. Xie}\thanks{xcxie@pku.edu.cn}
	\affiliation{International Center for Quantum Materials, School of Physics, Peking University, Beijing 100871, China}
	\affiliation{Beijing Academy of Quantum Information Sciences, West Bld.3,
		No.10 Xibeiwang East Rd., Haidian District, Beijing 100193, China}
	\affiliation{CAS Center for Excellence in Topological Quantum Computation,
		University of Chinese Academy of Sciences, Beijing 100190, China}
	\begin{abstract}
		Inside a three-dimensional strong topological insulator, a tube with $h/2e$ magnetic flux carries a pair of protected one-dimensional linear fermionic modes. This phenomenon is known as the ``wormhole effect". In this work, we find that the ``wormhole effect", as a unique degree of freedom, introduces exotic transport phenomena and thus manipulates the transport properties of topological insulators.
		Our numerical results demonstrate that the transport properties of a double-wormhole system can be manipulated by the wormhole interference. Specifically, the conductance and local density of states both oscillate with the Fermi energy due to the interference between the wormholes.
		Furthermore, by studying the multi-wormhole systems, we find that the number of wormholes can also modulate the differential conductance through a $\mathbb{Z}$$_{2}$ mechanism.
		Finally, we propose two types of topological devices in real applications, the ``wormhole switch" device and the ``traversable wormhole" device, which can be finely tuned by controlling the wormhole degree of freedom.
	\end{abstract}
	
	\maketitle
	
	
	\section{\label{sec:level1}Introduction}
	
	Ever since their discovery, topological insulators (TIs) have provided versatile
	platforms for physicists and material scientists to investigate nontrivial properties of the matter \cite{RevModPhys.82.3045,RevModPhys.83.1057}. A three-dimensional (3D) TI is a band insulator inside the bulk but carries metallic surface states protected by the time-reversal symmetry \cite{PhysRevLett.98.106803,PhysRevB.76.045302,PhysRevB.75.121306,chen_analytis_chu_liu_mo_qi_zhang_lu_dai_fang_etal_2009,zhang_liu_qi_dai_fang_zhang_2009,xia_qian_hsieh_wray_pal_lin_bansil_grauer_hor_cava_etal_2009}. The existence of these nontrivial surface states makes TIs ideal candidates for designing low-dissipation electronic devices. Recently, novel physical properties of the surface states have intrigued great interests among physicists. Especially, theoretical studies demonstrated that there is no gapless surface state in a TI nanowire because of the spin-momentum locking induced $\pi$ Berry phase \cite{PhysRevLett.105.206601,PhysRevLett.105.136403}.
	By applying a $\pi$ (in units of $\hbar/e$) magnetic flux, the effect of the $\pi$ Berry phase is eliminated, making the spectrum of the surface states gapless again. Furthermore, an $h/e$ period Aharonov-Bohm oscillation of the magneto-conductance was observed in different TI nanowires \cite{hong_zhang_cha_qi_cui_2014,cho_dellabetta_zhong_schneeloch_liu_gu_gilbert_mason_2015,wang_li_yu_liao_2016,PhysRevB.95.235436,PhysRevB.100.235307}.

	Interestingly, a flux tube inside the TI bulk shows the same energy spectrum as the TI nanowire \cite{PhysRevB.87.205409}. The $\pi$-flux tube carries a pair of gapless linear modes, providing a conducting channel for electrons to tunnel between the opposite surfaces of the TI. Thus the surface electrons that are spatially separated faraway can be connected by such a conducting flux tube. This phenomenon is called the ``wormhole effect" and the conducting flux tube is named as the ``wormhole" \cite{PhysRevB.82.041104}, for it acts like a wormhole in general relativity that bridges surfaces separated by an insulating bulk. Previously, by varying the magnetic flux $\phi$, numerical studies revealed the existence of the $\pi$-flux wormhole and its evolution \cite{PhysRevB.87.205409,PhysRevB.82.041104}.
	Such a $\phi$-dependence nature inspires us to study the exotic transport phenomena and the corresponding topological device applications based on the wormhole effect. However, systematic transport simulations of the wormhole systems were not yet performed. The main difficulty is that the existing numerical methods rely on the diagonalization of the 3D bulk Hamiltonian, which severely reduces the computation speed and limits studies to very small systems only \cite{PhysRevB.87.205409,PhysRevB.82.041104}. For this reason, we study the transport properties of the wormhole system in a new approach to overcome the difficulty caused by the bulk Hamiltonian.
	
	In this paper, we develop the model of the wormhole system based on a 2D effective lattice Hamiltonian (2DELH) for the surface of 3D strong TI, and study the transport properties of the system. This 2DELH captures the key physical properties of the topologically protected surface states without any reference to the bulk Hamiltonian. Therefore, it improves the computational speed significantly and makes it possible to study larger size and more complicated TI systems, especially the wormhole systems. Then, the transport properties of the wormhole system are investigated systematically based on the non-equilibrium Green's function method \cite{SPJ}.
	Our numerical results of the wormhole systems under different conditions demonstrate that the ``wormhole effect", as a unique degree of freedom, brings exotic transport phenomena and makes it feasible to manipulate the transport properties of TIs.
	By studying a double-wormhole system, we find the differential conductance and the local density of states (LDOS) oscillate with the Fermi energy due to interference between the wormholes. Therefore, the transport properties of TIs can be manipulated through the wormhole interference. Then, in multi-wormhole systems, we find that the number of wormholes can also modulate the differential conductance. These phenomena can be explained by an inter-wormhole backscattering mechanism. Particularly, in the perspective of transport, this inter-wormhole backscattering mechanism can be directly related to the $\mathbb{Z}$$_{2}$ classification in TIs. In applications, we propose two topological devices that are highly tunable by controlling the wormhole degree of freedom. One is the ``wormhole switch" device, where the conducting and insulating status of the device can be switched by varying the magnetic flux among the wormholes. The other is the ``traversable wormhole" device, in which a pair of conducting wormholes serves as a bridge that connects isolated regions, and enables electrons to travel between them.
	
	This paper is organized as follows. In Sec.~\ref{Model and methods}, we introduce the 2DELH for the wormhole systems and the numerical methods to calculate the transport observables. In Sec.~\ref{Interfering transport between wormholes}, we investigate the interfering transport between wormholes for the double- and the multi-wormhole systems. In Sec.~\ref{Topological devices based on the wormhole effect}, we propose two topological device applications based on the wormhole systems. Finally, we conclude the paper with a perspective in Sec.~\ref{Conclusion and perspectives}.
	\section{\label{Model and methods}Model and methods}
	\begin{figure}[t]
		\includegraphics[width=1\linewidth]{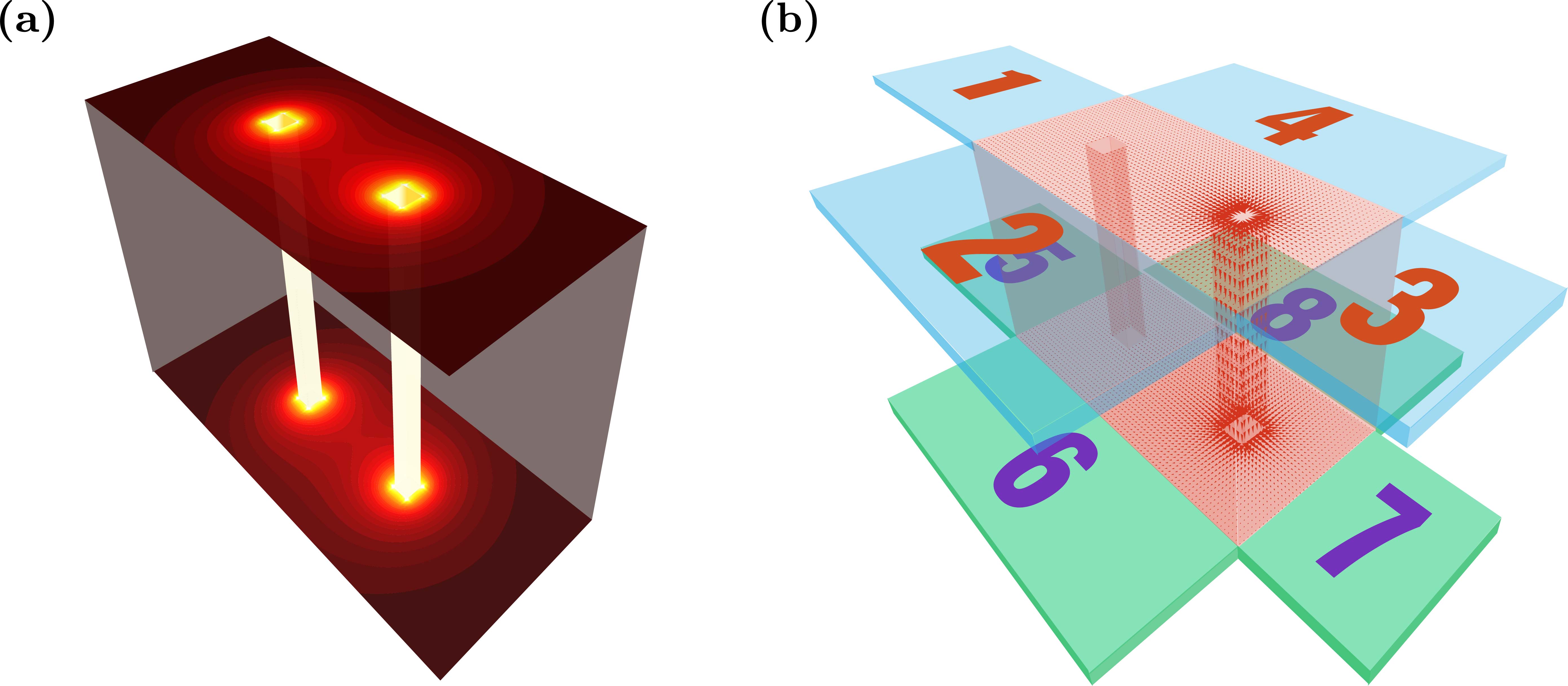}
		\caption{\label{fig:1}(a) Sketch of a double-wormhole system. Here, the flux tubes (with light color) inside the TI bulk represent the wormholes. (b) Schematic diagram of the studied multi-terminal device. Each terminal contact with the double-wormhole central region is considered as a source or drain. Electrons can transport between these terminals.}
	\end{figure}
	\subsection{\label{sec:level2}Surface Hamiltonian and 2D lattice model of the wormhole system}
	We establish the lattice model for the wormhole system based on the 2DELH, which incorporates a so-called Wilson mass term to overcome the fermion doubling problem \cite{PhysRevB.95.245137,NIELSEN198120,NIELSEN1981219,RevModPhys.55.775,PhysRevB.86.155146}. The low energy effective Hamiltonian of this model is:
	\begin{eqnarray}
	H(\mathbf{k})=\hbar v_{F}(\hat{\sigma} \times \mathbf{k}) \cdot \hat{n}+\frac{Wa}{2} k_{n}^{2} \sigma_{z}
	\label{eq:one},
	\end{eqnarray}
	where $v_{F}$ is the Fermi velocity of the topological surface states and $\hat{\sigma} \equiv\left(\sigma_{x}, \sigma_{y}, \sigma_{z}\right)$ with $\sigma_{x, y, z}$ the Pauli matrices, $\mathbf{k}=\left(k_{x}, k_{y}, k_{z}\right)$ is the wave vector, and $\hat{n}$ is the outward normal vector of the surface. The second term in $H(\mathbf{k})$ is the Wilson mass term and $k_{n}$ is the component of the wave vector perpendicular to $\hat{n}$ direction. In $x$-$y$ plane, the discretized 2D square-lattice Hamiltonian is written as:
	\begin{eqnarray}
	H_{2D}=&&\sum_{\mathbf{i}} \frac{i \hbar v_{F}}{2 a}(c_{\mathbf{i}}^{\dagger} \sigma_{y} c_{\mathbf{i}+\delta \hat{\mathbf{x}}}-c_{\mathbf{i}}^{\dagger} \sigma_{x} c_{\mathbf{i}+\delta \hat{\mathbf{y}}}) \nonumber\\
	&&-\sum_{\mathbf{i}}\frac{W}{2a}(c_{\mathbf{i}}^{\dagger} \sigma_{z} c_{\mathbf{i}+\delta \hat{\mathbf{x}}}+c_{\mathbf{i}}^{\dagger} \sigma_{z} c_{\mathbf{i}+\delta \hat{\mathbf{y}}})+\mathrm{H.c.}\nonumber\\
	&&+\sum_{\mathbf{i}}\frac{2W}{a}c_{\mathbf{i}}^{\dagger} \sigma_{z} c_{\mathbf{i}}
	\label{eq:two},
	\end{eqnarray}
	where $c_{\mathbf{i}}$ and $c_{\mathbf{i}}^{\dagger}$ are the annihilation and creation operators on site $\mathbf{i}$. $\delta \hat{\mathbf{x}}~(\delta \hat{\mathbf{y}})$ is the primitive vector of the square lattice along the $x (y)$ direction, and $a$ is the lattice constant. Although the incorporation of the Wilson mass term successfully overcomes the Fermi doubling problem by opening a gap at the redundant Dirac cones, it slightly shifts the Berry phase of the eigenstates $\psi_{\pm}(\mathbf{k})$ around the Fermi surface from $\gamma_{\pm}=\pi$ to $\gamma_{\pm}=\pi\left(1 \pm \frac{W a}{2 \hbar v_{F}} k_{F}\right)\approx\pi\left(1 \pm \frac{W a}{2 (\hbar v_{F})^{2}} E_{F}\right)$
	\cite{PhysRevB.95.245137,PhysRevLett.107.076801}. For this reason, the Fermi energy $E_{F}$ cannot be too large to hold the $\pi$-Berry phase. In this article, we take $a=1$, $\hbar v_{F}=1$ and $W=0.3$. Under this condition, the relative error in the Berry phase will be smaller than 3\% in the energy range -0.2\textless$E_{F}$\textless0.2. In order to include the magnetic flux, an additional phase factor $e^{i \phi_{ij}}$ is multiplied to the hopping term in Eq.(\ref{eq:two}). The summation of $\phi_{ij}$ along any closed loop, both around the wormhole and the surface, equals to the total magnetic flux inside it \cite{PhysRev.126.1636}.  With the help of Eq.(\ref{eq:one}) or (\ref{eq:two}), we can study the topological surface states determined by the 3D Hamiltonian within the 2D frame \cite{zhang_liu_qi_dai_fang_zhang_2009}, thus can greatly enhance the computational efficiency.
	
	Nevertheless, it is far from enough to construct the Hamiltonian of the wormhole system with Eq.(\ref{eq:one}) or Eq.(\ref{eq:two}) only. As discussed in Ref.~\cite{PhysRevLett.103.196804} and \cite{JPSJ.82.074712}, the low energy effective Hamiltonian with the form like Eq.(\ref{eq:one}) only works on a flat surface. On a curved surface (e.g. see FIG.~\ref{fig:1}), the curvature enters the Hamiltonian through the non-Abelian spin connection. Thus, the discretized lattice Hamiltonian can no longer be written in a concise form as Eq.(\ref{eq:two}).
	To solve such a problem, we propose a simple method in Appendix \ref{A} to deal with the discretized 2D Dirac Hamiltonian on the curved surface. The spirit of this method is dividing the whole surface of the wormhole system into a set of flat surfaces, on which the 2D Dirac Hamiltonians have the form of Eq.(\ref{eq:one}) or Eq.(\ref{eq:two}). Then, the surface of the wormhole system is reconstructed by gluing these flat surfaces together through unitary transformations on a locally defined spinor basis. The validity of the 2DELH is verified in Appendix \ref{B} by investigating the single-wormhole system.

	\subsection{Multi-terminal system and transport calculation method}
	
	Our numerical calculations of the differential conductance $G$, LDOS $\rho \left(E, r_{i}\right)$ and local current distribution $J_{i \rightarrow j}$ are based on the non-equilibrium Green's function method. As sketched in FIG.~\ref{fig:1} (b), the studied wormhole system is viewed as an eight-terminal device. Each terminal is considered as a source or drain electrode. Electrons can propagate between these terminals \footnote{Here, we ignore the side surfaces during the calculation. Generally, the mobility of the TI material is low. Therefore, in large systems, when electrodes are placed on the top and bottom surfaces, the electric current flowing through the side surface is negligible.}. The electric current $I_{n}$ for terminal $n$ is obtained by the Landauer-Büttiker formula \cite{datta_2007,haug_jauho_2010}:
	\begin{eqnarray}
	I_{n}=\left(e^{2}/h\right)\sum_{m\neq n}T_{nm}(V_{n}-V_{m})
	\label{eq:three},
	\end{eqnarray}
	where the transmission coefficient between terminal $n$ and $m$ is $T_{nm}$=$\operatorname{Tr}(\Gamma_{n}G^{r}\Gamma_{m}G^{a})$ with $G^{r/a}$ the retarded/advance Green's function of the central region. $\Gamma_{m/n}$=$i(\Sigma^{r}_{m/n}$$-$$\Sigma^{a}_{m/n})$ is the linewidth function at the $m/n$ terminal with $\Sigma_{m/n}^{r/a}$ the corresponding retarded/advanced self energy. During the calculation, we apply a small bias $V$ between these terminals (e.g. $V_{m}=0$, $V_{n}=V$). Then the total current $I$ can be obtained by summing all $I_{n}$s of the source terminals. The differential conductance $G$ is obtained by $G=dI/dV$.  LDOS at site $r_{i}$ can be expressed as:
	\begin{eqnarray}
	\rho \left(E, r_{i}\right)=-\frac{1}{\pi}\operatorname{Im} \operatorname{Tr}[G^{r}\left(E, r_{i}, r_{i}\right)]
	\label{eq:four},
	\end{eqnarray}
	where the trace is taken in the spin space and $E$ is the energy. Local current from site $i$ to site $j$ is \cite{PhysRevB.80.165316,PhysRevB.78.155413,PhysRevLett.87.126801}:
	\begin{eqnarray}
	J_{i\rightarrow j}&&=\frac{2e}{h}\sum_{\alpha,\beta}\sum_{n}\int_{-\infty}^{E-e V_{n}} \mathrm{d}E^{\prime}\operatorname{Im}\{H_{i\alpha,j\beta}[G^{r} \Gamma_{n} G^{a}]_{j\beta,i\alpha}\}\nonumber\\&&\approx\frac{2e}{h}\sum_{\alpha,\beta}\int_{-\infty}^{E} \mathrm{d}E^{\prime}\operatorname{Im}\{H_{i\alpha,j\beta}[G^{r} \sum_{n}\Gamma_{n} G^{a}]_{j\beta,i \alpha}\}\nonumber\\&&\quad-\frac{2 e^{2}}{h} \sum_{\alpha,\beta}\operatorname{Im}[H_{i\alpha,j\beta}\sum_{n}G_{j\beta,i\alpha}^{n}(E)V_{n}]
	\label{eq:five}.
	\end{eqnarray}
	$V_{n}$ denotes the applied bias on terminal $n$ and $G^{n}(E)=G^{r}(E) \Gamma_{n}(E) G^{a}(E)$. We only focus on the non-equilibrium current which corresponds to the second term in Eq.(\ref{eq:five}):
	\begin{eqnarray}
	J_{i\rightarrow j}^{\operatorname{non-eq}}&&=-\frac{2 e^{2}}{h} \sum_{\alpha,\beta}\operatorname{Im}[H_{i\alpha,j\beta}\sum_{n}G_{j\beta,i\alpha}^{n}(E)V_{n}]
	\label{eq:six}.
	\end{eqnarray}
	
	In order to clearly study the physical behavior of the wormholes, the influence of the finite-size effect of the top and bottom surfaces should be eliminated. The size of the surfaces should be taken as large as possible. Unfortunately, previous 3D Hamiltonian can only deal with small systems \cite{PhysRevB.82.041104,PhysRevB.87.205409,cho_dellabetta_zhong_schneeloch_liu_gu_gilbert_mason_2015}. The computation advantage of our 2D model enables us to study a substantial large system (about 200$\times$200 for a single-wormhole system and 300$\times$300 for a double-wormhole system). Besides, we shift the Fermi energy $E_{F}$ at each terminal by 0.3. In this way, the ohmic contact between the metallic electrodes and the wormhole system becomes perfect.
	
	\section{\label{Interfering transport between wormholes}Interfering transport between wormholes}
	
	\subsection{Double-wormhole interference}
	
	In this subsection, we demonstrate that the wormhole degree of freedom can be manipulated through the interfering transport in a double-wormhole system. Here, we treat terminals on the top or bottom surface as a whole. Bias $V$ is applied between source (terminals 1-4, $V_{1}$=$V_{2}$=$V_{3}$=$V_{4}$=$V$) and drain (terminals 5-8, $V_{5}$=$V_{6}$=$V_{7}$=$V_{8}$=$0$), as shown in FIG.~\ref{fig:1} (b). The geometry of the wormhole, for simplicity, is taken as a cuboid.
	
\begin{figure}[t]
		\includegraphics[width=1\linewidth]{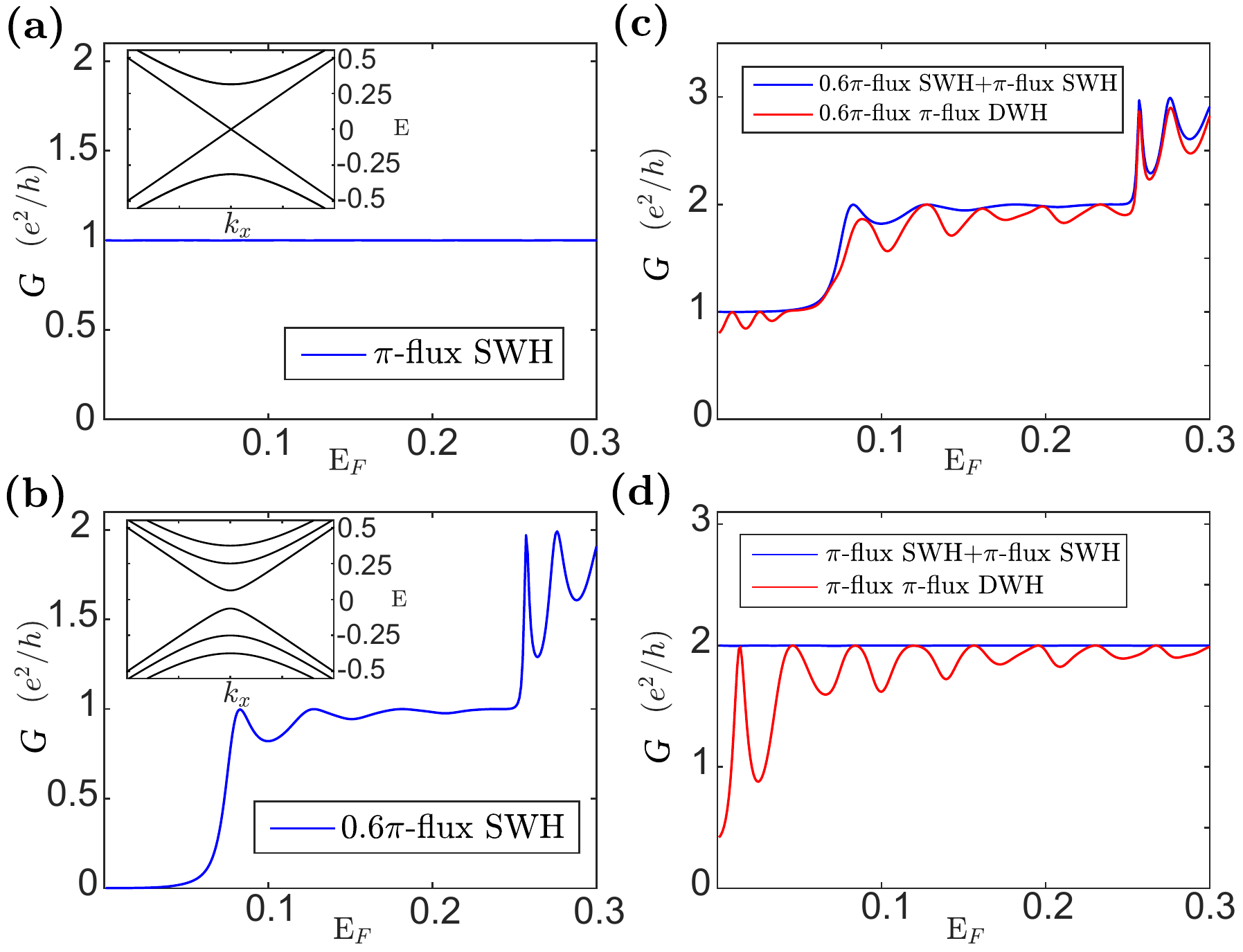}
		\caption{\label{fig:2} (a), (b) Differential conductance $G$ versus the Fermi energy $E_{F}$ in a single-wormhole (SWH) system with magnetic flux $\pi$ and 0.6$\pi$. Inset figures show the band structures of the corresponding infinitly long wormholes. (c), (d) Comparison between the single-wormhole and double-wormhole (DWH) systems in $G$. Distance between the two wormholes $D$=80. In each subfigure, the blue curve is obtained by directly summing over two differential conductance of the corresponding single-wormhole systems. The red curve shows $G$ of the double-wormhole system. The difference between the red and blue curves indicates a wormhole interference. Here, the cross section size of the wormhole in all subfigures is 11$\times$11, and the length $L$=100.}
	\end{figure}

	We first calculate the band structure of an infinitely long wormhole to count how many conducting channels contribute to the differential conductance. When the magnetic flux $\phi=\pi$ (in units of $\frac{\hbar}{e}$), the wormhole is gapless and carries a pair of 1D linear modes [see insets in FIG.~\ref{fig:2} (a)]. In contrast, when $\phi$ is shifted from $\pi$, as shown in FIG.~\ref{fig:2} (b), a gap is opened. The linear mode of the $\pi$-flux wormhole provides a conducting channel between the top and bottom surfaces, so that in the single-wormhole system, the differential conductance $G \equiv e^{2}/h$. Because of the gap, $G$ for the 0.6$\pi$-flux single-wormhole system starts from 0 and then oscillates toward an integer conductance [FIG.~\ref{fig:2} (b)]. In a double-wormhole system, both wormholes contribute conducting channels. If the two wormholes are independent, i.e. there is no interference between them, the total differential conductance will be a naive summation $G=G_{1}+G_{2}$ [blue curves in FIG.~\ref{fig:2} (c) and (d)]. $G_{1,2}$ is the differential conductance of the corresponding single-wormhole system. However, our numerical results demonstrate that $G$ oscillates with the Fermi energy [the red curves in FIG.~\ref{fig:2} (c) and (d)], which implies that there is an interference between the two wormholes.
	
\begin{figure}[t]
		\includegraphics[width=1.0\linewidth]{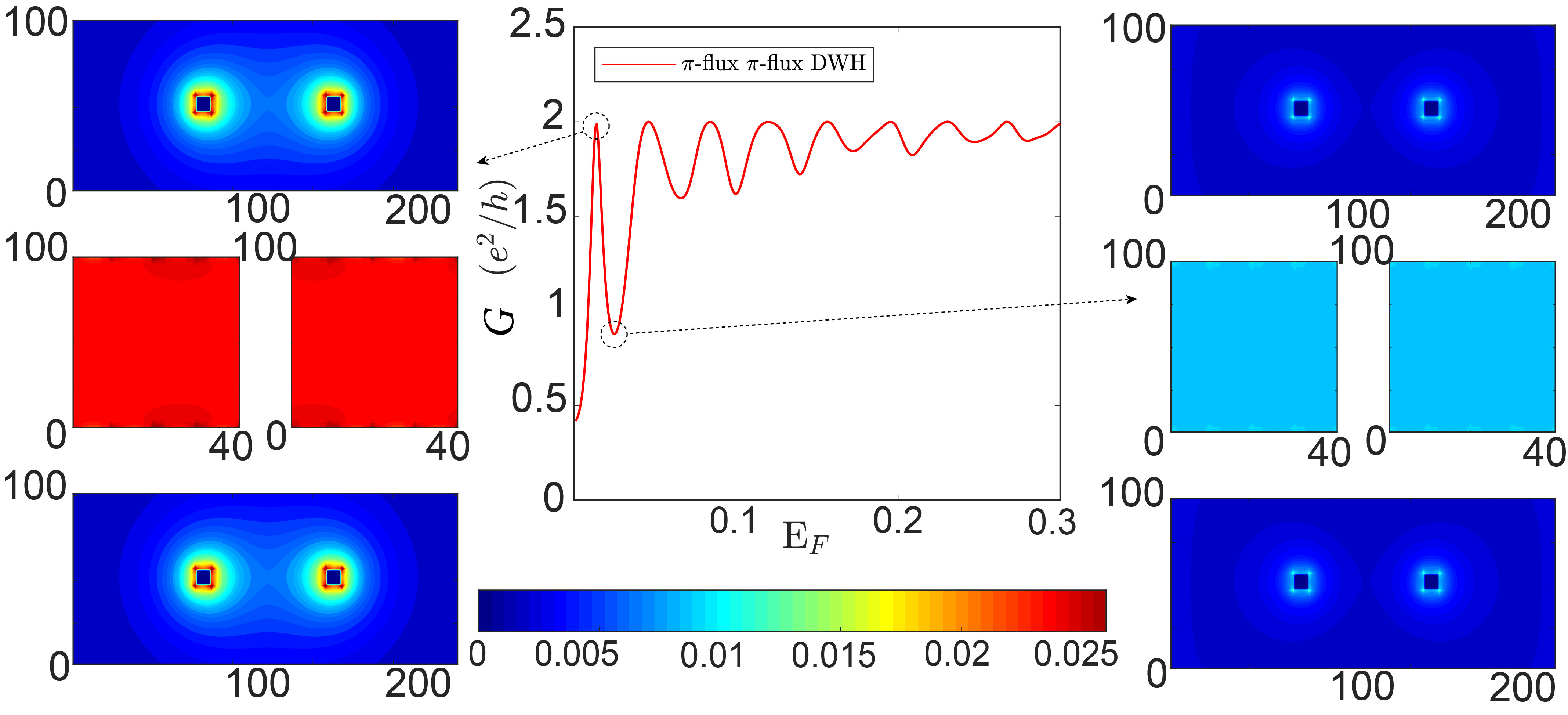}
		\caption{\label{fig:3} LDOS distribution $\rho(E_{F},r)$ of the double-wormhole system with two $\pi$-flux wormholes. The $G$-$E_{F}$ curve takes from FIG.~\ref{fig:2} (f). The LDOS  distribution $\rho(E_{F},r)$ at the first peak and dip on the $G$-$E_{F}$ curve are plotted correspondingly. In each subplot, the top and bottom panels correspond to the top and bottom surfaces. The two small panels in the middle represent the expanded side surfaces of the two wormholes.}
	\end{figure}

	In order to see the details of the interference, we calculate the LDOS distributions $\rho(E_{F},r)$ of the double-wormhole system (see FIG.~\ref{fig:3}). In the left and right panels, the LDOS distribution at the first peak and dip on the $G$-$E_{F}$ curve are plotted. The oscillation of $G$ is intimately related to the oscillation of $\rho$. For example, $\rho$ in the wormholes is much larger at the peak than at the dip of $G$. The rising of $\rho$ also indicates that there is resonant tunneling between the top and bottom surfaces modulated by the wormhole interference.
	
	We further investigate the geometrical dependence of double-wormhole interference. In the following, we compare the differential conductance $G$ by varying the distance $D$, length $L$, and the cross-section size of the wormholes.
	
\begin{figure}[b]
		\includegraphics[width=1.02\linewidth]{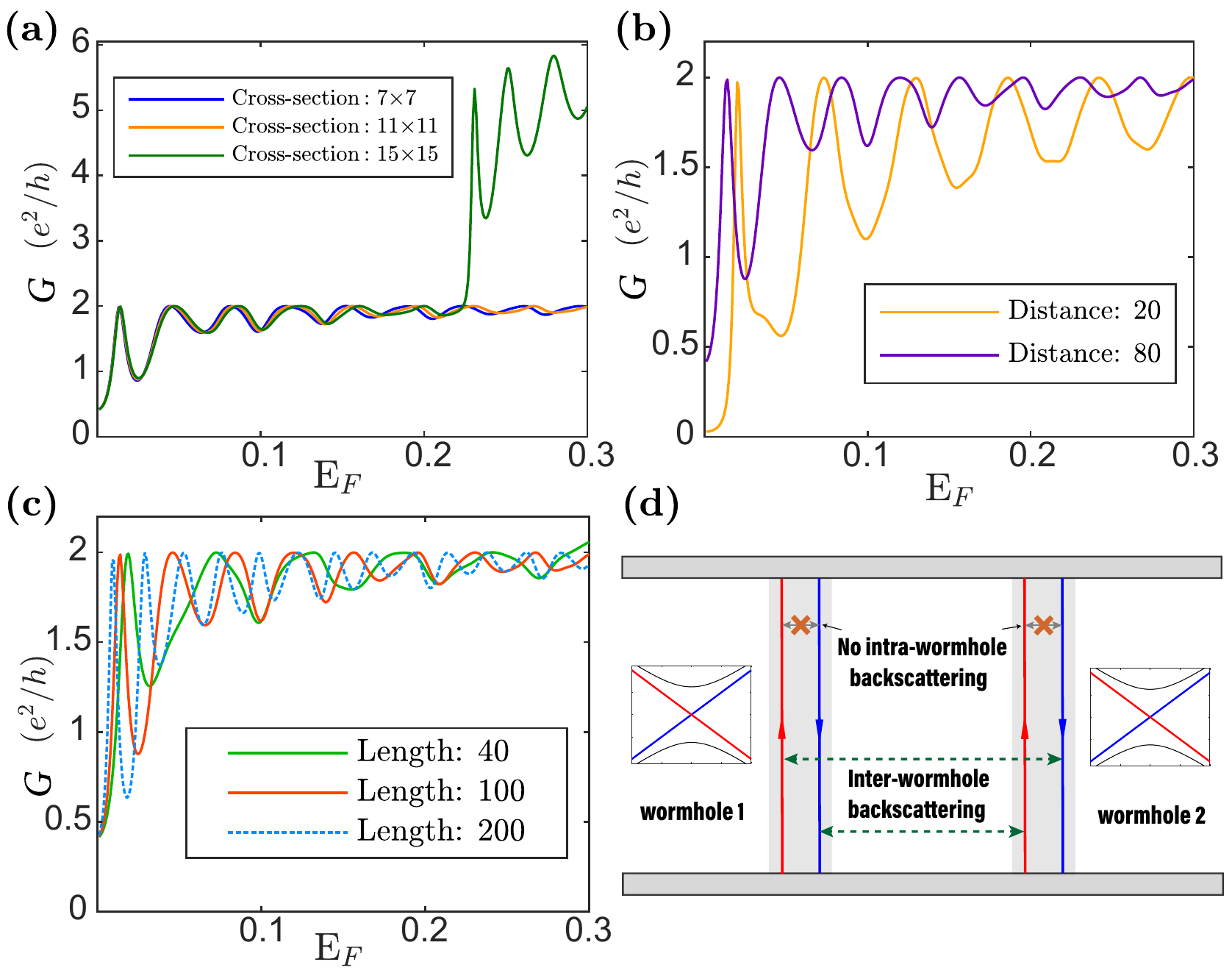}
		\caption{\label{fig:4} Geometrical dependence of the double-wormhole interference. (a) The differential conductance $G$ of double-wormhole systems with different cross-section size. The length of the wormhole $L$=100 and the distance between them $D$=80. (b) $G$ versus $E_{F}$ under different $D$ with $L$=100.
			(c) $G$ versus $E_{F}$ by varying $L$, with fixed $D$=80. The cross-section size in (b) and (c) takes 11$\times$11. (d) Schematic of the inter-wormhole backscattering process between two $\pi$-flux wormholes. Red(blue) arrowed lines inside the wormholes represent upward(downward) linear modes. Backscattering of linear modes can(cannot) happen between inter(intra)-wormholes.}
	\end{figure}

	As shown in FIG.~\ref{fig:4} (a), $G$ with different cross-section sizes almost coincide with each other when E$_{F}$\textless0.2. Therefore the cross-section sizes of wormholes has almost no influence on the double-wormhole interference. Nevertheless, the cross-section size still plays a role in tuning the wormhole degree of freedom. The green curve jumps when E$_{F}$\textgreater0.2, because a wider wormhole provides more confined subbands (for a cylindrical TI nanowire with radius $R$, the energy difference between confined subband $\propto$ $1/R$ \cite{PhysRevB.82.041104}). The Fermi energy $E_{F}$ crosses three subbands (including two degenerate subbands) for each wormhole. There are six conducting channels, and $G$ reaches 6$e^{2}/h$.
	
	FIG.~\ref{fig:4} (b) and (c) show the behavior of the wormhole interference with fixed cross-section size (11$\times$11) by varying $D$ and $L$. As plotted in FIG.~\ref{fig:4} (b), the interference becomes stronger when the two wormholes get closer, so that the amplitude of the oscillation in $G$ is enhanced. Meanwhile, enlarging $L$ and $D$ can visibly suppress the oscillation period [see FIG.~\ref{fig:4} (b) and (c)]. The $L$ and $D$ dependence of the oscillation can be phenomenologically explained by considering the double-wormhole system as an effective Aharonov-Bohm ring \cite{PhysRev.115.485,schuster_buks_heiblum_mahalu_umansky_shtrikman_1997}, where the two wormholes together with the top and bottom surfaces form a closed path, and the total perimeter is $2(L+D)$. The resonance tunneling happens when $2k_{F}(L+D)=2\pi(n+1/2)$, where $k_{F}$ is the wave vector at $E_{F}$, $n$ is an integer, and $1/2$ originates from the geometrical phase of the linear band. The oscillation period $\Delta E$ equals to the energy difference between neighbor resonance peaks, thus $\Delta E=\hbar v_{F}\Delta k_{F}=2\pi \hbar v_{F}/2(L+D)$. Therefore, the oscillation of $G$ is faster for the double-wormhole system with longer wormhole length $L$ or distance $D$.

	In order to better understand the interference phenomenon in double-wormhole systems and the formation of the effective AB ring, we propose an inter-wormhole backscattering mechanism, as sketched in FIG.~\ref{fig:4} (d). The $\pi$-flux wormhole carries a pair of gapless linear modes that are protected by the topological bulk and the induced magnetic flux $\phi$. Due to the Klein tunneling \cite{RN82}, there is no backscattering between the two linear modes in a $\pi$-flux wormhole, and $G\equiv e^{2}/h$ is observed [FIG.~\ref{fig:2} (c)]. When $\phi\neq\pi$, the two modes are coupled, and a gap is opened in the energy spectrum [see inset in FIG.~\ref{fig:2} (b)]. This coupling induces a backscattering between the two linear modes (intra-wormhole backscattering) and causes the oscillation of $G$ even in a single-wormhole system [see $G$-$E_{F}$ curve in FIG.~\ref{fig:2} (b)]. In a double-wormhole system with two $\pi$-flux wormholes, there is no intra-wormhole backscattering. However, linear modes of different wormholes are coupled with each other through the top and bottom surfaces. This coupling results in an inter-wormhole backscattering [see dashed arrows in FIG.~\ref{fig:4} (d)] and can also lead to the oscillation of $G$. The closer the wormholes are, the stronger the coupling between the backscattering channels is, thus the oscillation amplitude of $G$ is enhanced [FIG.~\ref{fig:4} (b)]. Furthermore, because of the inter-wormhole backscattering, electrons can travel between wormholes and their spatial trajectory form a closed loop [see FIG.~\ref{fig:4} (d)], thus the double-wormhole system can be viewed as an effective AB ring.
	
	\subsection{Multi-wormhole interference}
	\begin{figure}[b]
		\includegraphics[width=1\linewidth]{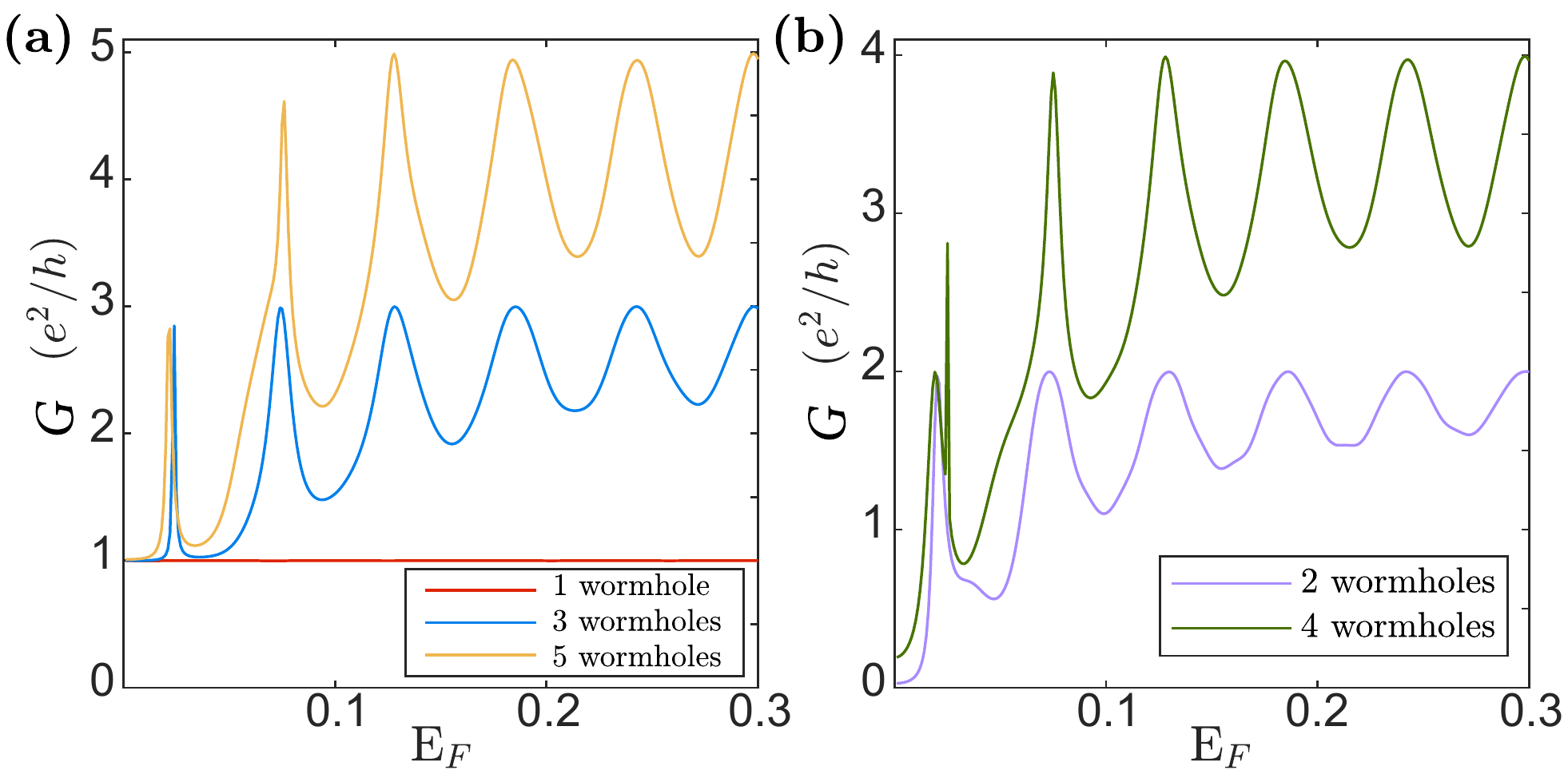}
		\caption{\label{fig:5} $G$-$E_{F}$ for multi-wormhole systems. Wormholes are lined up with distance $D$=20, and wormhole length $L$=100. (a) Multi-wormhole systems with 1,~3,~5 wormholes. $G \geqslant e^{2}/h$, indicating at least one conducting channel survives. (b) Multi-wormhole systems with 2,~4 wormholes. $G$ can be smaller than $e^{2}/h$.}
	\end{figure}
	
	In this subsection, we show that unique transport phenomena can be obtained through changing the number of wormholes. Particularly, we find that the differential conductance $G$ shows an parity dependence on the number of wormholes. It originates from the fact that the physics behind the inter-wormhole backscattering is the same as the backscattering between the helical edge modes in 2D TIs, and falls into a $\mathbb{Z}$$_{2}$ classification \cite{PhysRevB.79.241303}.
	
	On the edge of 2D TIs, the forward edge mode cannot be backscattered into its time-reversal counterpart (the corresponding helical counterpart), but can be backscattered into backward edge modes that belong to different Kramers pairs. This backscattering mechanism leads to a  $\mathbb{Z}$$_{2}$ classification for time-reversal invariant materials \cite{PhysRevLett.95.146802}. A system with an even (odd) number of helical edge mode Kramers pairs is categorized into a $\mathbb{Z}$$_{2}$=0 ($\mathbb{Z}$$_{2}$=1) class, which is equivalent to a normal (topological) insulator. In normal insulators, the backscattering can destroy all conducting channels, and the minimum of $G$ falls to 0. On the contrary, in TIs there is always one protected gapless edge mode survives the backscattering, thus $G \geqslant e^{2}/h$.
	
	Interestingly, the multi-wormhole system (all wormholes are $\pi$-flux) reproduces this $\mathbb{Z}$$_{2}$ classification through the wormhole backscattering. A pair of linear modes in each wormhole can be considered as a Kramers pair of helical edge modes, and the intra (inter)-wormhole backscattering is forbidden (allowed). In analogy with the 2D TI, multi-wormhole systems with odd (even) number of wormholes are expected to be in the $\mathbb{Z}$$_{2}$=1 ($\mathbb{Z}$$_{2}$=0) or ``topological" (``normal") classification.
	Our numerical simulations of multi-wormhole systems with 1,3,5 and 2,4 wormholes strongly confirm these expectations [see FIG.~\ref{fig:5} (a) and (b)]. For multi-wormhole systems with an odd number of wormholes, $G \geqslant e^{2}/h$ is observed [FIG.~\ref{fig:5} (a)]. It is consistent with the ``topological" case ($\mathbb{Z}$$_{2}$=1) where one protected mode survives the inter-wormhole backscattering.
	By contrast, as shown in FIG.~\ref{fig:5} (b), $G$-$E_{F}$ curves of systems with an even number of wormholes start from $G\approx0$. It belongs to the  ``normal" case ($\mathbb{Z}$$_{2}$=0), where no conducting mode survives the inter-wormhole backscattering.
	The modulation of $G$ through changing the number of wormholes provides a new way to manipulate the wormhole degree of freedom, enabling wormhole systems to be more controllable in topological device applications.

	\section{\label{Topological devices based on the wormhole effect}Topological devices based on the wormhole effect}
	In this section, we show that the manipulation of the wormhole degree of freedom makes the wormhole systems feasible in designing novel topological devices. Specifically, we propose two types of devices: (1) the ``wormhole switch", which controls the transport behavior of the device by varying the magnetic flux $\phi$; (2) the ``traversable wormhole" device, which provides conducting channels and enables electrons to travel across the surface separated by obstacles.
	\begin{figure}[t]
		\includegraphics[width=1\linewidth]{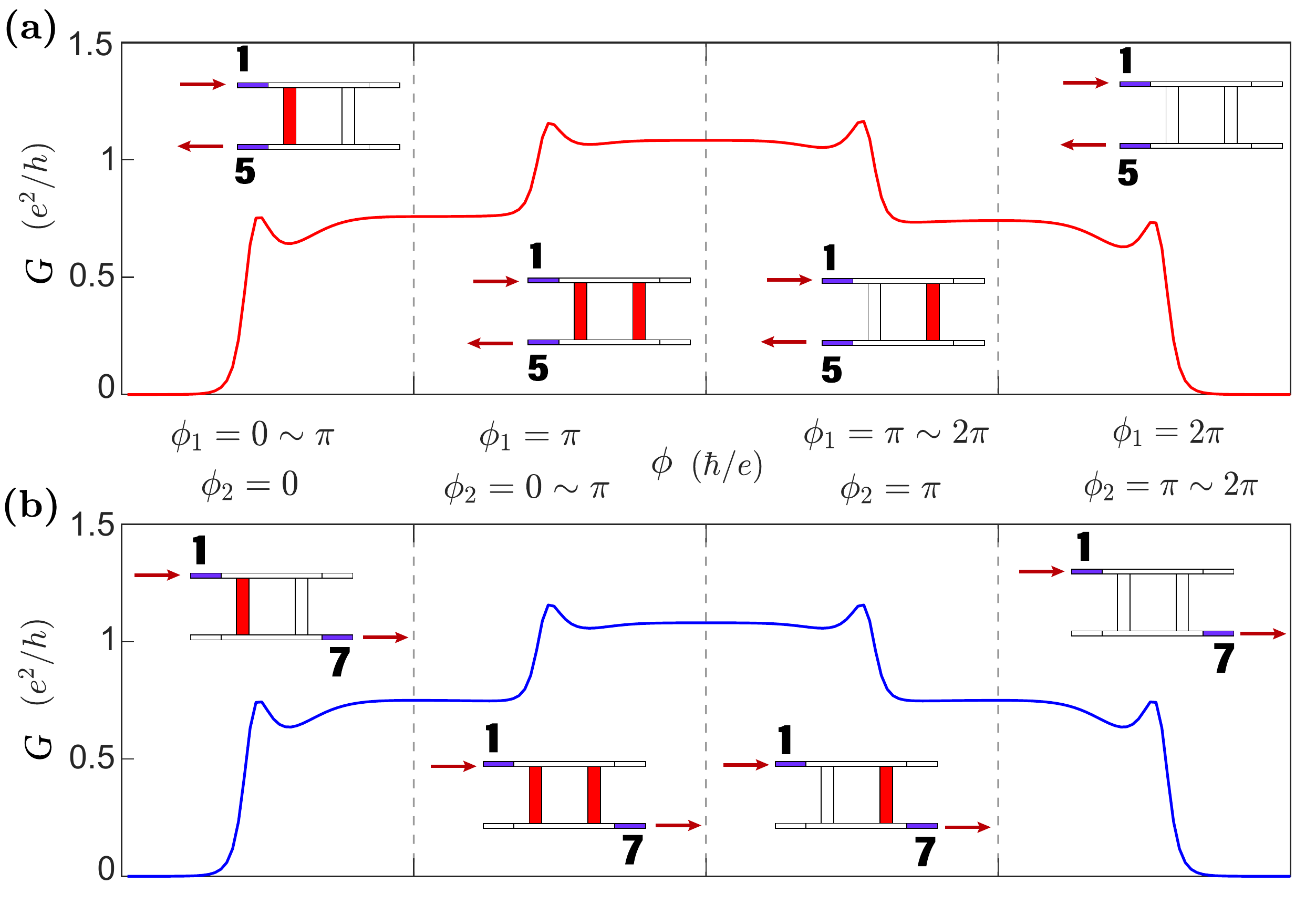}
		\caption{\label{fig:6} $G$ of ``wormhole switch" device by alternately varying $\phi$ in the two wormholes.  Here, $E_{F}$=0.1, $D$=80, $L$=100 and the cross-section size takes 11$\times$11. In each inset, the red (white) -colored area represents a conducting (insulating) wormhole. Terminal 1(5) in (a) and terminal 1(7) in(b) are taken as source (drain) with all the other terminals set float.}
	\end{figure}
	\begin{figure}[t]
		\includegraphics[width=1\linewidth]{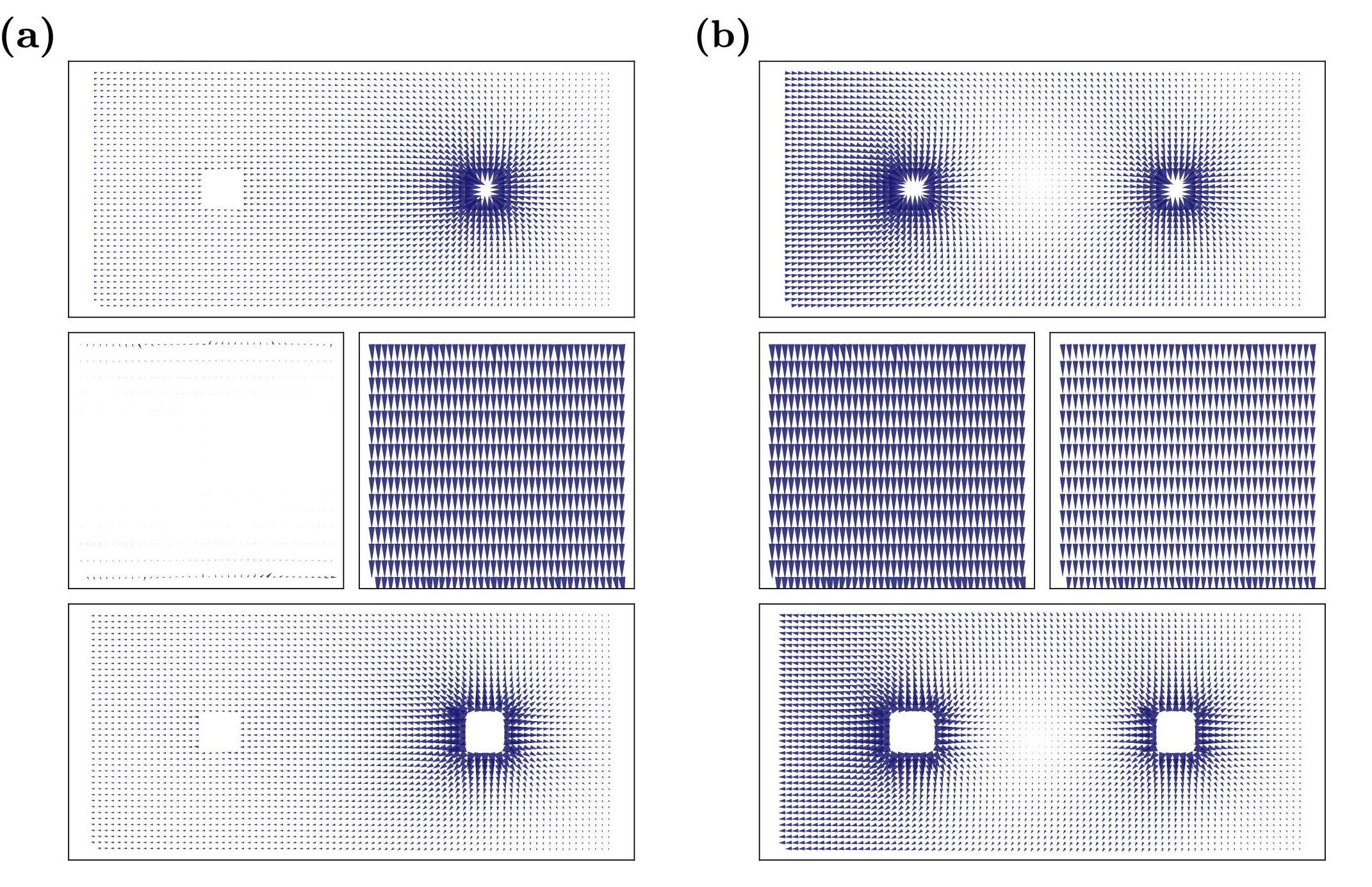}
		\caption{\label{fig:7} Local current distribution of the ``wormhole switch" device. In both subfigures, the upper and lower panels correspond to the top and bottom surfaces. The middle two panels correspond to the expanded side surfaces of the wormholes. Terminals 1 and 5 are chosen as the source and drain. Device size and $E_{F}$ are the same as FIG.~\ref{fig:6} (a). (a) $\phi_{1}$=$0$ and $\phi_{2}$=$\pi$. (b) $\phi_{1}$=$\phi_{2}$=$\pi$.}
	\end{figure}
	\subsection{``Wormhole switch" device}
	
	By manipulating the wormhole degree of freedom through varying $\phi$, the conducting and insulating status of wormholes can be switched. This phenomenon inspires us to utilize wormhole systems as switches in topological devices. In this subsection, we study such an application by investigating the $\phi$-dependence of double-wormhole systems.
	
	Here we pick two of the eight terminals as external leads and set all the other terminals open. Two cases are considered, where the electric current comes from lead 1 and leaves at lead 5 or lead 7 [see insets of  FIG.~\ref{fig:6}]. FIG.~\ref{fig:6} (a) and (b) plot the differential conductance $G$ versus $\phi$ in the two wormholes. For both cases, $G$=0 in the absence of $\phi$, signaling the insulating status between the top and bottom surfaces. By alternately varying $\phi$ of the two wormholes, $G$ jumps steeply and then manifests a plateau-like behavior. This phenomenon means that the system can be steadily switched between the conducting (on) and insulating (off) status. In fact, we can not only switch the on and off status of the topological device but also switch the spatial distribution of the current on each wormhole. By calculating the local current distribution (here, terminals 1 and 5 are chosen as the source and drain), we find a ``remote tunneling" phenomenon, as shown in FIG.~\ref{fig:7} (a). The near wormhole is ``turned off", and the distant wormhole is ``turned on". The electric current coming from terminal 1 bypassing the near wormhole, then flows through the distant conducting wormhole to terminal 5. For comparison, FIG.~\ref{fig:7} (b) shows the case where both wormholes are ``turned on". More electric current flows through the near wormhole (about 55\%) than through the distant wormhole (about 45\%).
	In applications, the controllable of the current distribution enables us to integrate complex topological devices, which can realize various transport functions among different places.
	
	More figuratively, the wormhole provides a ``bridge" that connects opposite surfaces of the TI and makes electrons traveling between them possible. Thus the wormhole in TIs resembles the real wormhole in general relativity. The difference is that in TIs, the wormhole degree of freedom is highly tunable by varying the magnetic fluxes. In applications, we can switch not only the on and off status, but also the current distribution of the topological device.

	\subsection{``Traversable wormhole" device}
	\begin{figure*}
		\includegraphics[width=0.85\linewidth]{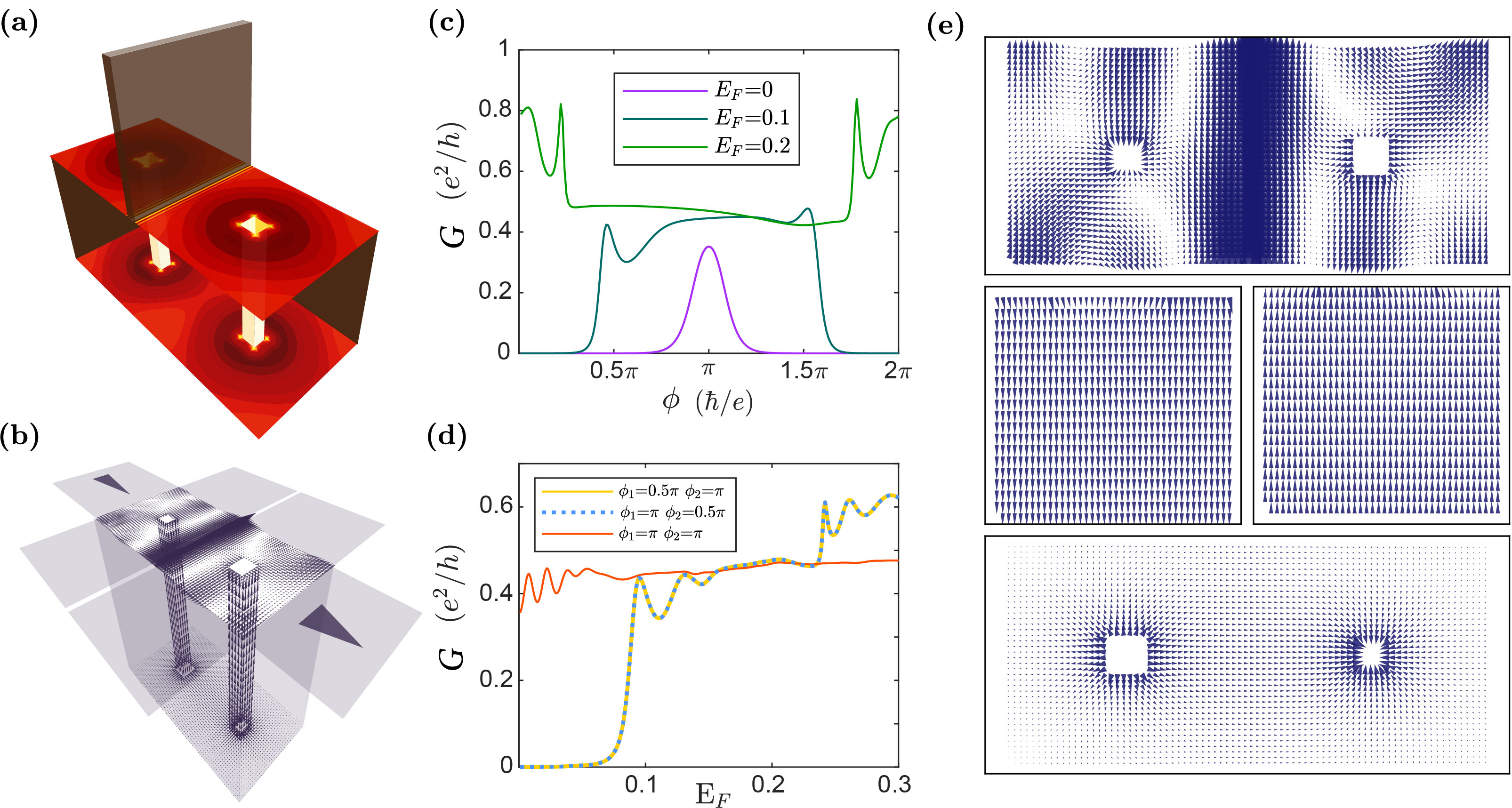}
		\caption{\label{fig:8}  (a) Sketch of a ``traversable wormhole" device. (b) Schematic of the current distribution in the ``traversable wormhole" device. Electric current flows from terminal 1 traveling across the two conducting wormholes, then to terminal 3. The separated regions in the top surface of TI are connected by wormholes. (c) $G$ as a function of the magnetic flux $\phi$ at different $E_{F}$. $\phi_{1}$=$\phi_{2}$=$\phi$. (d) $G$ as a function of $E_{F}$. Three different combinations of $\phi_{1}$ and $\phi_{2}$ are compared. (e) Local current distribution of the ``traversable wormhole" device with $\phi_{1}$=$\phi_{2}$=$\pi$ and $E_{F}$=$0.1$. The top surface is separated in the middle, the conducting wormhole pair serves as a bridge that connects the separated top surface. Device size is the same as FIG.~\ref{fig:6}.}
	\end{figure*}
	Apart from connecting the opposite surfaces isolated by the insulating bulk, the wormhole in TIs can also provide conducting channels to connect the separated regions of the same surface. In real topological devices, the transport of electrons on the same surface of TI is not always unimpeded. For example, the metallic surface state may be destroyed by magnetic impurities, or separated by lumps, external leads and other complicated device structures [e.g., see FIG.~\ref{fig:8} (a)]. Thus, electrons cannot pass through these obstacles. In order to make the electronic transport between the separated regions possible, similar as the space traveling between separated spacetime \cite{PhysRev.48.73}, we propose a ``traversable wormhole" device. The obstacles on the surface of TI are modeled by a cutting off in the middle on the top surface so that electrons cannot transport across it directly [see FIG.~\ref{fig:8} (b)]. The local current distribution [see FIG.~\ref{fig:8} (e)] shows clearly that electric current comes from the source (terminal 1) flows to the bottom surface through the near wormhole, then flows back to the drain on the top surface (terminal 3) through the distant wormhole, completes a ``wormhole traveling".
	
	Furthermore, we investigate the transport performance of the device in detail.
	As shown in FIG.~\ref{fig:8} (c), the differential conductance $G$ can be modulated by simultaneously varying the magnetic fluxes of both wormholes with a fixed $E_{F}$. We find that the conducting region falls into a flux window centered at $\pi$. The jump of $G$ from 0 to a finite value means that the separate regions are connected by wormholes. The width of the flux window is enlarged with the increasing of $E_{F}$, signaling the tunability of the ``traversable wormhole" device.
	Then we compare $G$ under three combinations of the magnetic fluxes $\phi_{1}$ and $\phi_{2}$ [FIG.~\ref{fig:8} (d)]. The ``wormhole traveling" phenomenon happens at any $E_{F}$ only if both wormholes are $\pi$-flux. This is because only when $E_{F}$ crosses the subbands of both wormholes [see insets FIG.~\ref{fig:2} (a) and (b)] can the two wormholes provide a conducting channel for the electron to complete a ``wormhole traveling" process.
	
	In applications, the ``traversable wormhole" device can bridge TI surfaces that are separated by obstacles. For example, when a TI thin film is fabricated on the substrate, the bottom surface of the thin film is sealed and maintains excellent transport performance. However, its top surface is exposed to the environment or be used to build complex device structures. Thus the electron transport on the top surface may be impeded. Fortunately, with the help of the ``traversable wormhole" device, electrons can still travel across the top surface. Moreover, through varying $E_{F}$ or $\phi$ of the wormholes, the traveling can be finely controlled.

	\section{\label{Conclusion and perspectives}Conclusion and perspective}
	
	In summary, with the help of the developed model based on the 2DELH, we have numerically studied the wormhole systems. We find that the ``wormhole effect", as a unique degree of freedom, can manipulate the transport properties of TIs.
	The oscillation of the differential conductance $G$ and the LDOS in the double-wormhole system demonstrated an interfering transport between the wormholes, depending on the system geometry. The interference phenomenon originates from the inter-wormhole backscattering. Electrons can be backscattered between the wormholes to complete the motion in a closed path, causing an effective AB interference. Furthermore, by studying the multi-wormhole systems, we find that the inter-wormhole backscattering mechanism leads to a $\mathbb{Z}$$_{2}$ classification, and $G$ shows an parity dependence on the number of wormholes. Therefore, the transport properties of the system can be modulated by the wormhole numbers.
	
	We then propose two topological device applications through the manipulation of the wormhole degree of freedom. The first one is the ``wormhole switch" device, which enables us to switch not only the on and off status, but also the current distribution of the device through varying the magnetic fluxes of the wormholes. The second one is the ``traversable wormhole" device, which provides conducting channels to connect isolated regions of the topological surfaces. It enables electrons to complete a ``wormhole traveling" and bypass the obstacles on the surface.
	
	Recently, transport experiments based on semiconductor, 2D TI quantum well, or superconductor thin films with nanopore structures have been reported \cite{maier,du_wang,yang_liu}. We expect that the TI thin film with nanopore structures can serve as an ideal platform to realize the wormhole devices. Since the radius of the etched nanopores is reported to be around 50$\sim$100 nm, the applied magnetic field required for a flux quantum is expected to be less than 0.25 T. Therefore, the wormhole devices and the manipulations of the wormhole degree of freedom are highly achievable in experiments. Other potential candidates to realize the wormhole systems are the artificial structures such as the photonic/phononic crystals \cite{RevModPhys.91.015006,lu_joannopoulos_2014,liu_chen_xu_2019,PhysRevApplied.5.054021,cai_ye_qiu_xiao_yu_ke_liu_2020}
	and the topological electric circuit \cite{PhysRevX.5.021031,PhysRevLett.114.173902}, which have attracted extensive attention. Especially, the 3D photonic topological insulator \cite{yang_gao_xue_} and the 3D acoustic TI \cite{PhysRevLett.123.195503} have been realized recently. The macroscopic scale of these structures makes it easier to build the tubular geometry of the wormholes. Moreover, introducing and controlling the magnetic fluxes are feasible in classical wave systems \cite{PhysRevX.5.021031,RN55,RN32,RN68,RN59,RN60}. For these reasons, artificial structures provide flexible platforms for physicists to investigate the novel properties of the wormhole systems.

	\begin{acknowledgments}
		We thank Yan-Feng Zhou and Huai-Ming Guo for fruitful discussion. This work is financially supported by the National Basic Research Program of China (Grants No.~2017YFA0303301, and No.~2019YFA0308403) and the National Natural Science
		Foundation of China (Grants No.~11534001, No.~11674028, No.~11822407, and No.~11921005).
	\end{acknowledgments}

	\appendix
	\section{Construction of the 2D effective lattice Hamiltonian for the wormhole system}
	\label{A}
	\begin{figure*}
		\includegraphics[width=1\linewidth]{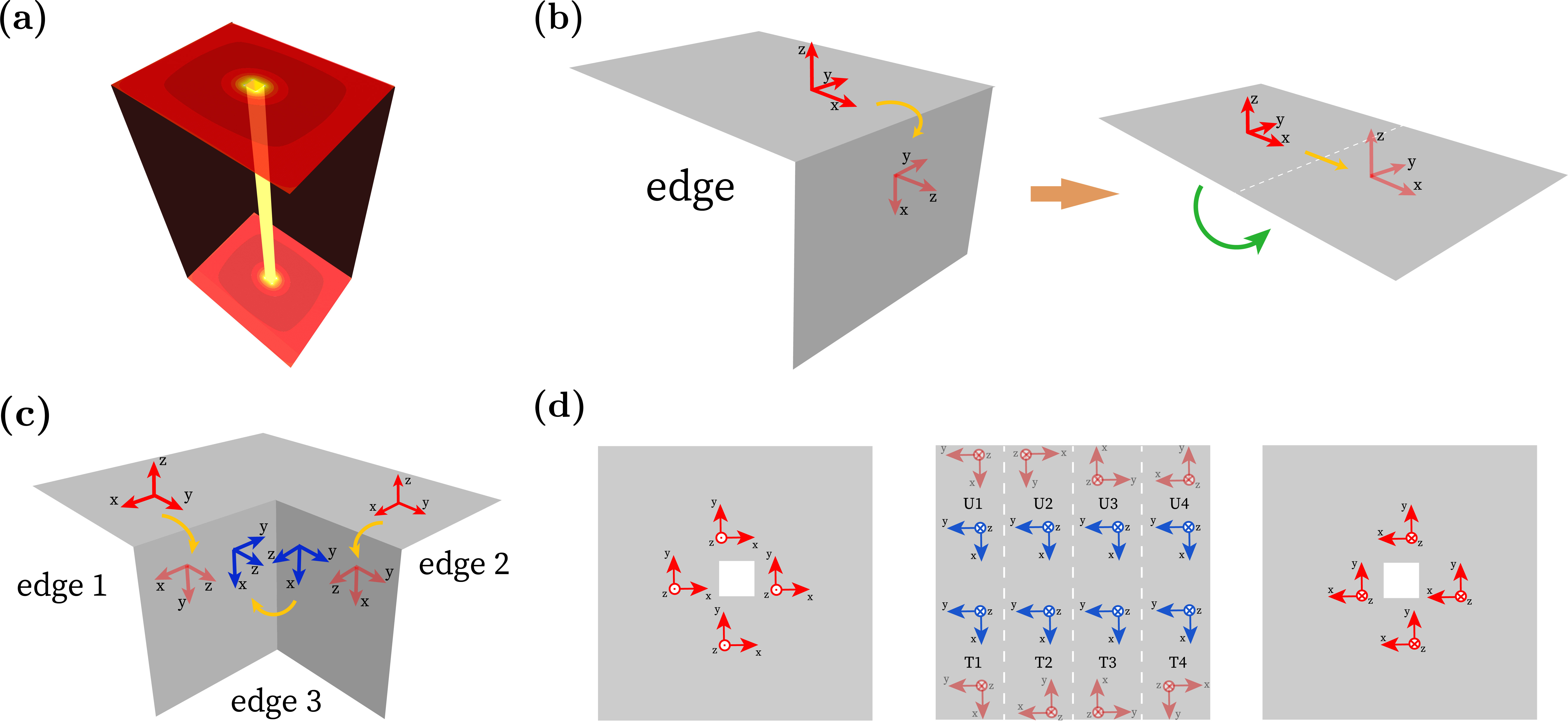}  
		\caption{\label{fig:9} (a) Sketch of a single-wormhole system. (b) A bent surface with one edge can be flattened. The local coordinate frame is parallelly translated along the surface.  (c) A corner-like surface can be viewed as three sets of bent surfaces. The red(blue)-colored coordinate frame represents the local coordinate frame on the top(side) surface. The light red colored ones represent the coordinate frames parallelly translated from the top surface. (d) Coordinate frame specified on the top surface (left panel), the surface of the wormhole (middle panel) and bottom surface (right panel). The z-direction is the outward normal direction on each surface (here the outward means the axis points from the bulk to the vacuum). Red(blue) colored frames represent local coordinate frames on the top and bottom surface (on the wormhole). Light-red colored frames are obtained by parallelly translating frames from top and bottom surfaces to the wormhole. The four slices of the wormhole surface are connected to the top surface in order of left-down-right-up. $U_{i}$ and $T_{i}$ indicate the unitary transformations of the corresponding spinor basis on the edges.}
	\end{figure*}
	
	As have been discussed in Ref.~\cite{PhysRevLett.103.196804} and \cite{JPSJ.82.074712}, the Dirac Hamiltonian $H(\mathbf{r})$ on a 2D curved surface is not trivial, since $H$ inherits geometrical and topological information of the surface. It brings difficulties in numerical calculations when $H$ is discretized into a lattice form. In this appendix, we propose a simple method to deal with the discretized $H(\mathbf{r})$ on a curved surface by dividing the whole surface into a set of flat surfaces and glue them together by unitary transformations on spinor basis at the edges. Then we give the exact form of the transformation matrices. This method can also be generalized to higher dimensions.
	
	In order to encode geometric information into the Hamiltonian $H(\mathbf{r})$, partial derivatives $\partial_{k}$ in the momentum term should be replaced by covariant derivatives \cite{PhysRevLett.103.196804}, which generates the parallel translations of the spinor basis on the curved surface. As shown in FIG.~\ref{fig:9} (b), a bent surface can be flattened because of its zero curvature ($\Omega$=0). The local coordinate frame on the flat surface can be parallelly translated without the additional information of $\Omega$. This is why the surface of an infinite long TI tube can be flattened \cite{PhysRevB.95.245137,PhysRevB.89.085305}, except for an anti-periodic boundary condition that we will discuss soon.
	
	In contrast, for curved surfaces with non-zero $\Omega$ such as the corner-like surface plotted in FIG.~\ref{fig:9} (c) (the cross point of the three edges is a singular point of $\Omega$), the flattening procedure can no longer be performed. Consequently, the parallel translations of local spinor basis cannot be unified on the three edges (they ``frustrate" each other), and at least two of the frames should be related by a nontrivial coordinate transformation [coordinate frames on the 1-3 surface (determined by edge 1 and edge 3) in FIG.~\ref{fig:9} (c)]. In the wormhole system, corner-like structures appear when the wormhole crosses the top and bottom surfaces. The planar graph of a single-wormhole system with specified local coordinate frames is shown in FIG.~\ref{fig:9} (d) (colored red for the top and bottom surfaces and blue for the wormhole). The light red colored coordinate frames are parallelly translated from top and bottom surfaces to the wormhole. Obviously, these frames are not always coincident with the blue colored frames, and coordinate transformations are inevitable.
	
	When $H(\mathbf{r})$ is discretized into a lattice form, the parallel translation operations are reflected in the hopping terms.
	To be clear, we consider an electron hopping process from the top surface of the corner-like structure in FIG.~\ref{fig:9} (c), to the 1-3 side surface. The wave function components after hopping are obtained by projecting the electron state onto the blue colored spinor basis on the wormhole, instead of the parallelly translated spinor basis (light red colored). Thus, wave functions before and after hopping should be related by a unitary transformation.

	For the wormhole system, as shown in FIG.~\ref{fig:9} (d), parallel translation of spinor basis between different coordinate frame induces a set of unitary transformations between the wormhole and the top (bottom) surface denoted by $U_{i}$ ($T_{i}$) respectively, where $i=1,2,3,4$.
	They can be obtained as follows.
	Suppose the local spinor basis on the top surface is $\ket{\mathbf{e}^{t}_{\mu}}$ ($\mu=\uparrow,\downarrow$ represents the $+z$ or $-z$ direction, similarly hereinafter), the four side surfaces of the wormhole are denoted by a,b,c,d on which the local spinor basis are $\ket{\mathbf{a}^{w}_{\mu}}$,$\ket{\mathbf{b}^{w}_{\mu}}$,$\ket{\mathbf{c}^{w}_{\mu}}$ and $\ket{\mathbf{d}^{w}_{\mu}}$. The parallelly translated spinor basis from the top surface to the side surfaces of the wormhole are denoted by $\ket{\mathbf{a}^{t}_{\mu}}$,$\ket{\mathbf{b}^{t}_{\mu}}$,$\ket{\mathbf{c}^{t}_{\mu}}$ and $\ket{\mathbf{d}^{t}_{\mu}}$.  For example, when an electron state $\ket{\psi}=\sum_{\mu}\ket{\mathbf{e}^{t}_{\mu}}v^{\mu}$ on the top surface is parallelly translated to the side surface b, the translated state is $\ket{\psi^{t}}=\sum_{\mu}\ket{\mathbf{b}^{t}_{\mu}}v^{\mu}$. Notice that in  numerical calculations, the wave functions after hopping is expressed under the $\ket{\mathbf{b}^{w}_{\mu}}$ basis, thus the unitary transformation from $\ket{\mathbf{b}^{t}_{\mu}}$ to $\ket{\mathbf{b}^{w}_{\mu}}$ is needed. They can be obtained by noticing that $\ket{\psi^{t}}=\sum_{\mu}\ket{\mathbf{b}^{t}_{\mu}}v^{\mu}=\sum_{\mu}\sum_{\nu}\ket{\mathbf{b}^{w}_{\nu}}\braket*{\mathbf{b}^{w}_{\nu}}{\mathbf{b}^{t}_{\mu}}v^{\mu}=\sum_{\mu}\sum_{\nu}\ket{\mathbf{b}^{w}_{\nu}}U_{\nu \mu}v^{\mu}$ and the transformation matrix is $U_{\nu \mu}=\braket*{\mathbf{b}^{w}_{\nu}}{\mathbf{b}^{t}_{\mu}}$. Importantly, the parallelly translated basis $\ket{\mathbf{b}^{t}_{\mu}}$ can be obtained by rotating the original basis $\ket{\mathbf{e}^{t}_{\mu}}$ and we have $\ket{\mathbf{b}^{t}_{\mu}}=\mathbf{\hat R}_{b\leftarrow e}\ket{\mathbf{e}^{t}_{\mu}}$, $\mathbf{\hat R}_{b\leftarrow e}$ is a rotation operator.
	Similarly, $\ket{\mathbf{a}^{w}_{\mu}}=\mathbf{\hat R}_{a\leftarrow e}\ket{\mathbf{e}^{t}_{\mu}}$ and $\ket{\mathbf{b}^{w}_{\mu}}=\mathbf{\hat R}_{b\leftarrow a}\ket{\mathbf{a}^{w}_{\mu}}$.
	From these relations we can write down the matrix form of $U$ as:
	\begin{eqnarray}
	\nonumber
	U_{\mu \nu}&&=\braket*{\mathbf{b}^{w}_{\mu}}{\mathbf{b}^{t}_{\nu}}\\
	&&=\bra{\mathbf{b}^{w}_{\mu}}\mathbf{\hat R}_{b\leftarrow e}\mathbf{\hat R}_{a\leftarrow e}^{\dagger}\mathbf{\hat R}_{b\leftarrow a}^{\dagger}\ket{\mathbf{b}^{w}_{\nu}}
	\label{A1}.
	\end{eqnarray}
	\begin{figure*}
		\includegraphics[width=1.0\linewidth]{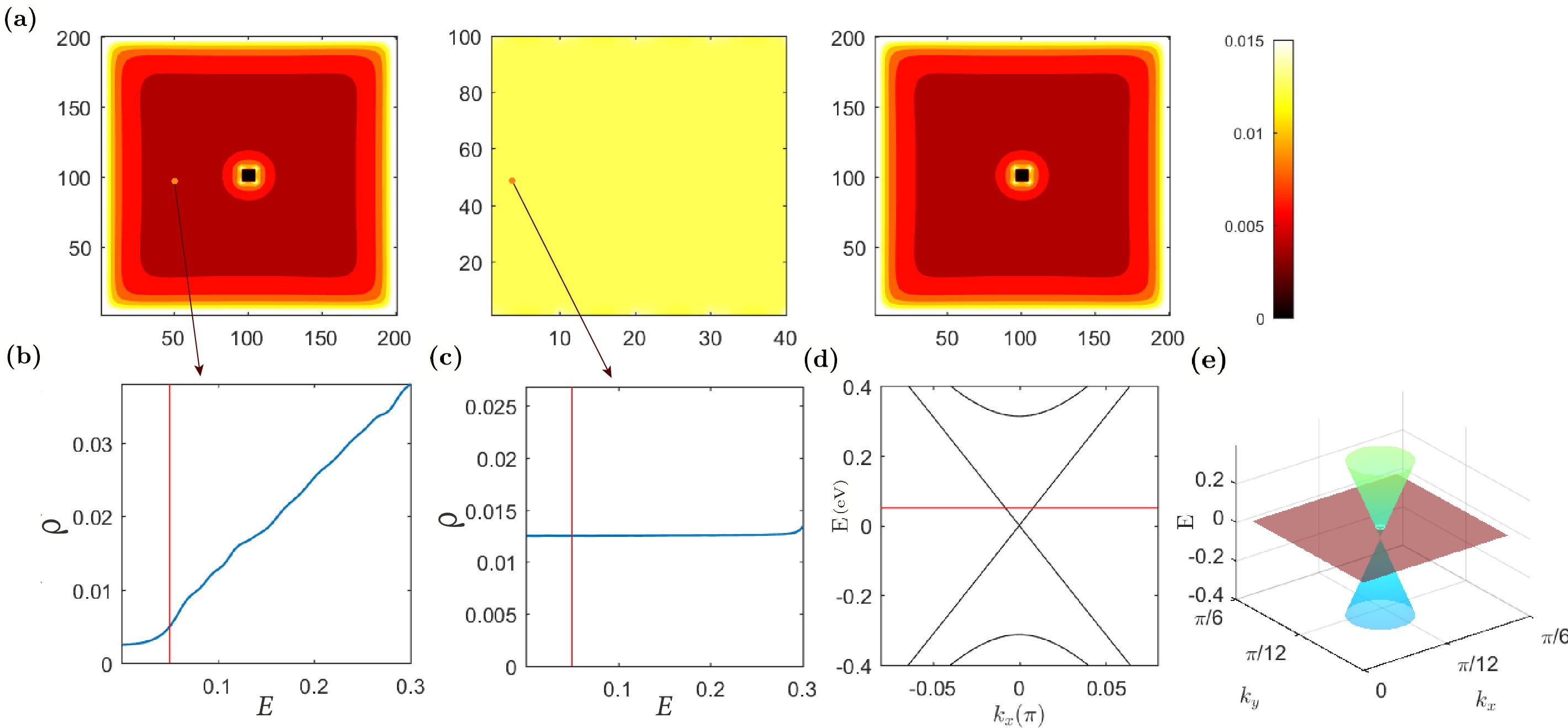}
		\caption{\label{fig:10} (a) LDOS $\rho$ of the single-wormhole system at Fermi energy $E_{F}=0.05$. (b) $\rho$-$E$ at a fixed point on the surface. (c) $\rho$-$E$ at a fixed point on the wormhole. (e) Energy dispersion of an infinitely long wormhole. (f) Energy dispersion of a TI surface. The cross-section size of the wormhole takes 11$\times$11, the wormhole length $L$=100. The red lines(surface) in (b)-(e) represent the Fermi energy.}
	\end{figure*}
	From Eq.(\ref{A1}) we can calculate the exact form of the unitary transformation between the top surface and the side surface b on the wormhole. Express $\mathbf{\hat R}$ under the $\ket{\mathbf{b}^{w}_{\mu}}$ basis (for example, the matrix form of $\mathbf{\hat R}_{b\leftarrow e}$ rotation is $e^{-i\frac{\sigma_{y}}{4}\pi}$), we have:
	\begin{eqnarray}
	U_{2}=U_{y}(\frac{\pi}{2})U_{z}(-\frac{\pi}{2})U_{x}(\frac{\pi}{2})=U_{z}(-\frac{\pi}{2})
	\label{A2},
	\end{eqnarray}
	where $U_{x,y,z}(\phi)$=$e^{-i\frac{\sigma_{x,y,z}}{2}\phi}$. After similar analysis, we can also obtain the unitary transformations between the top surface and the a,c,d side surfaces on the wormhole, with:
	\begin{eqnarray}
	&&U_{1}=I_{2 \times 2}\nonumber\\
	&&U_{3}=U_{y}(\frac{\pi}{2})U_{y}(\frac{\pi}{2})U_{x}(\pi)=U_{z}(-\pi)\nonumber\\
	&&U_{4}=U_{y}(\frac{\pi}{2})U_{z}(\frac{\pi}{2})U_{x}(\frac{3\pi}{2})=U_{z}(-\frac{3\pi}{2})
	\label{A3},
	\end{eqnarray}
	Then, the unitary transformations between the bottom surface and the wormhole are:
	\begin{eqnarray}
	&&T_{1}=I_{2 \times 2}\nonumber\\
	&&T_{2}=U_{x}(\frac{\pi}{2})U_{z}(-\frac{\pi}{2})U_{y}(-\frac{\pi}{2})=U_{z}(-\frac{\pi}{2})\nonumber\\
	&&T_{3}=U_{y}(-\frac{\pi}{2})U_{z}(-\pi)U_{y}(-\frac{\pi}{2})=U_{z}(-\pi)\nonumber\\
	&&T_{4}=U_{x}(-\frac{\pi}{2})U_{z}(-\frac{3\pi}{2})U_{y}(-\frac{\pi}{2})=U_{z}(-\frac{3\pi}{2})
	\label{A4}.
	\end{eqnarray}
	
	For the last step, the flat surface should be rolled up into a tube and forms the wormhole. The anti-periodic boundary condition should be applied \cite{PhysRevB.95.245137}, because after a 2$\pi$ rotation, the spinor basis changes by a $\pi$ phase factor. Therefore the unitary transformation, in this case, is just $-\mathbf{I}$. Physically it can also be understood as a $\pi$ geometrical phase induced by the rotation of spinors \cite{book:826289}.

	\section{Properties of single-wormhole systems}
	\label{B}
	
	In this appendix, we investigate the properties of a single-wormhole system and check the validity of the 2DELH. The corresponding LDOS of a $\pi$-flux wormhole system is studied [FIG.~\ref{fig:10} (a)]. Figures (d) and (e) plot the energy dispersions of the infinite long wormhole and the infinite large surface. Figures (b) and (c) show the LDOS at fixed points on the top surface and on the wormhole. As expected, the total density of states $\rho$ is a linear function of $E$ for the topological surface due to 2D Dirac cone dispersion, and a constant for the $\pi$-flux wormhole due to 1D linear dispersion. Our numerical results perfectly match this expectation, therefore establish our confidence for further calculations based on this 2DELH.

	\bibliography{draft}

\begin{thebibliography}{58}%
\makeatletter
\providecommand \@ifxundefined [1]{%
 \@ifx{#1\undefined}
}%
\providecommand \@ifnum [1]{%
 \ifnum #1\expandafter \@firstoftwo
 \else \expandafter \@secondoftwo
 \fi
}%
\providecommand \@ifx [1]{%
 \ifx #1\expandafter \@firstoftwo
 \else \expandafter \@secondoftwo
 \fi
}%
\providecommand \natexlab [1]{#1}%
\providecommand \enquote  [1]{``#1''}%
\providecommand \bibnamefont  [1]{#1}%
\providecommand \bibfnamefont [1]{#1}%
\providecommand \citenamefont [1]{#1}%
\providecommand \href@noop [0]{\@secondoftwo}%
\providecommand \href [0]{\begingroup \@sanitize@url \@href}%
\providecommand \@href[1]{\@@startlink{#1}\@@href}%
\providecommand \@@href[1]{\endgroup#1\@@endlink}%
\providecommand \@sanitize@url [0]{\catcode `\\12\catcode `\$12\catcode
  `\&12\catcode `\#12\catcode `\^12\catcode `\_12\catcode `\%12\relax}%
\providecommand \@@startlink[1]{}%
\providecommand \@@endlink[0]{}%
\providecommand \url  [0]{\begingroup\@sanitize@url \@url }%
\providecommand \@url [1]{\endgroup\@href {#1}{\urlprefix }}%
\providecommand \urlprefix  [0]{URL }%
\providecommand \Eprint [0]{\href }%
\providecommand \doibase [0]{https://doi.org/}%
\providecommand \selectlanguage [0]{\@gobble}%
\providecommand \bibinfo  [0]{\@secondoftwo}%
\providecommand \bibfield  [0]{\@secondoftwo}%
\providecommand \translation [1]{[#1]}%
\providecommand \BibitemOpen [0]{}%
\providecommand \bibitemStop [0]{}%
\providecommand \bibitemNoStop [0]{.\EOS\space}%
\providecommand \EOS [0]{\spacefactor3000\relax}%
\providecommand \BibitemShut  [1]{\csname bibitem#1\endcsname}%
\let\auto@bib@innerbib\@empty
\bibitem [{\citenamefont {Hasan}\ and\ \citenamefont
  {Kane}(2010)}]{RevModPhys.82.3045}%
  \BibitemOpen
  \bibfield  {author} {\bibinfo {author} {\bibfnamefont {M.~Z.}\ \bibnamefont
  {Hasan}}\ and\ \bibinfo {author} {\bibfnamefont {C.~L.}\ \bibnamefont
  {Kane}},\ }\bibfield  {title} {\bibinfo {title} {Colloquium: Topological
  insulators},\ }\href {https://doi.org/10.1103/RevModPhys.82.3045} {\bibfield
  {journal} {\bibinfo  {journal} {Rev. Mod. Phys.}\ }\textbf {\bibinfo {volume}
  {82}},\ \bibinfo {pages} {3045} (\bibinfo {year} {2010})}\BibitemShut
  {NoStop}%
\bibitem [{\citenamefont {Qi}\ and\ \citenamefont
  {Zhang}(2011)}]{RevModPhys.83.1057}%
  \BibitemOpen
  \bibfield  {author} {\bibinfo {author} {\bibfnamefont {X.-L.}\ \bibnamefont
  {Qi}}\ and\ \bibinfo {author} {\bibfnamefont {S.-C.}\ \bibnamefont {Zhang}},\
  }\bibfield  {title} {\bibinfo {title} {Topological insulators and
  superconductors},\ }\href {https://doi.org/10.1103/RevModPhys.83.1057}
  {\bibfield  {journal} {\bibinfo  {journal} {Rev. Mod. Phys.}\ }\textbf
  {\bibinfo {volume} {83}},\ \bibinfo {pages} {1057} (\bibinfo {year}
  {2011})}\BibitemShut {NoStop}%
\bibitem [{\citenamefont {Fu}\ \emph {et~al.}(2007)\citenamefont {Fu},
  \citenamefont {Kane},\ and\ \citenamefont {Mele}}]{PhysRevLett.98.106803}%
  \BibitemOpen
  \bibfield  {author} {\bibinfo {author} {\bibfnamefont {L.}~\bibnamefont
  {Fu}}, \bibinfo {author} {\bibfnamefont {C.~L.}\ \bibnamefont {Kane}},\ and\
  \bibinfo {author} {\bibfnamefont {E.~J.}\ \bibnamefont {Mele}},\ }\bibfield
  {title} {\bibinfo {title} {Topological insulators in three dimensions},\
  }\href {https://doi.org/10.1103/PhysRevLett.98.106803} {\bibfield  {journal}
  {\bibinfo  {journal} {Phys. Rev. Lett.}\ }\textbf {\bibinfo {volume} {98}},\
  \bibinfo {pages} {106803} (\bibinfo {year} {2007})}\BibitemShut {NoStop}%
\bibitem [{\citenamefont {Fu}\ and\ \citenamefont
  {Kane}(2007)}]{PhysRevB.76.045302}%
  \BibitemOpen
  \bibfield  {author} {\bibinfo {author} {\bibfnamefont {L.}~\bibnamefont
  {Fu}}\ and\ \bibinfo {author} {\bibfnamefont {C.~L.}\ \bibnamefont {Kane}},\
  }\bibfield  {title} {\bibinfo {title} {Topological insulators with inversion
  symmetry},\ }\href {https://doi.org/10.1103/PhysRevB.76.045302} {\bibfield
  {journal} {\bibinfo  {journal} {Phys. Rev. B}\ }\textbf {\bibinfo {volume}
  {76}},\ \bibinfo {pages} {045302} (\bibinfo {year} {2007})}\BibitemShut
  {NoStop}%
\bibitem [{\citenamefont {Moore}\ and\ \citenamefont
  {Balents}(2007)}]{PhysRevB.75.121306}%
  \BibitemOpen
  \bibfield  {author} {\bibinfo {author} {\bibfnamefont {J.~E.}\ \bibnamefont
  {Moore}}\ and\ \bibinfo {author} {\bibfnamefont {L.}~\bibnamefont
  {Balents}},\ }\bibfield  {title} {\bibinfo {title} {Topological invariants of
  time-reversal-invariant band structures},\ }\href
  {https://doi.org/10.1103/PhysRevB.75.121306} {\bibfield  {journal} {\bibinfo
  {journal} {Phys. Rev. B}\ }\textbf {\bibinfo {volume} {75}},\ \bibinfo
  {pages} {121306} (\bibinfo {year} {2007})}\BibitemShut {NoStop}%
\bibitem [{\citenamefont {Chen}\ \emph {et~al.}(2009)\citenamefont {Chen},
  \citenamefont {Analytis}, \citenamefont {Chu}, \citenamefont {Liu},
  \citenamefont {Mo}, \citenamefont {Qi}, \citenamefont {Zhang}, \citenamefont
  {Lu}, \citenamefont {Dai}, \citenamefont {Fang},\ and\ \citenamefont
  {et~al.}}]{chen_analytis_chu_liu_mo_qi_zhang_lu_dai_fang_etal_2009}%
  \BibitemOpen
  \bibfield  {author} {\bibinfo {author} {\bibfnamefont {Y.~L.}\ \bibnamefont
  {Chen}}, \bibinfo {author} {\bibfnamefont {J.~G.}\ \bibnamefont {Analytis}},
  \bibinfo {author} {\bibfnamefont {J.-H.}\ \bibnamefont {Chu}}, \bibinfo
  {author} {\bibfnamefont {Z.~K.}\ \bibnamefont {Liu}}, \bibinfo {author}
  {\bibfnamefont {S.-K.}\ \bibnamefont {Mo}}, \bibinfo {author} {\bibfnamefont
  {X.~L.}\ \bibnamefont {Qi}}, \bibinfo {author} {\bibfnamefont {H.~J.}\
  \bibnamefont {Zhang}}, \bibinfo {author} {\bibfnamefont {D.~H.}\ \bibnamefont
  {Lu}}, \bibinfo {author} {\bibfnamefont {X.}~\bibnamefont {Dai}}, \bibinfo
  {author} {\bibfnamefont {Z.}~\bibnamefont {Fang}},\ and\ \bibinfo {author}
  {\bibnamefont {et~al.}},\ }\bibfield  {title} {\bibinfo {title} {Experimental
  realization of a three-dimensional topological insulator,
  $\mathrm{Bi_{2}Te_{3}}$},\ }\href {https://doi.org/10.1126/science.1173034}
  {\bibfield  {journal} {\bibinfo  {journal} {Science}\ }\textbf {\bibinfo
  {volume} {325}},\ \bibinfo {pages} {178–181} (\bibinfo {year}
  {2009})}\BibitemShut {NoStop}%
\bibitem [{\citenamefont {Zhang}\ \emph {et~al.}(2009)\citenamefont {Zhang},
  \citenamefont {Liu}, \citenamefont {Qi}, \citenamefont {Dai}, \citenamefont
  {Fang},\ and\ \citenamefont {Zhang}}]{zhang_liu_qi_dai_fang_zhang_2009}%
  \BibitemOpen
  \bibfield  {author} {\bibinfo {author} {\bibfnamefont {H.}~\bibnamefont
  {Zhang}}, \bibinfo {author} {\bibfnamefont {C.-X.}\ \bibnamefont {Liu}},
  \bibinfo {author} {\bibfnamefont {X.-L.}\ \bibnamefont {Qi}}, \bibinfo
  {author} {\bibfnamefont {X.}~\bibnamefont {Dai}}, \bibinfo {author}
  {\bibfnamefont {Z.}~\bibnamefont {Fang}},\ and\ \bibinfo {author}
  {\bibfnamefont {S.-C.}\ \bibnamefont {Zhang}},\ }\bibfield  {title} {\bibinfo
  {title} {Topological insulators in $\mathrm{Bi_{2}Se_{3}}$,
  $\mathrm{Bi_{2}Te_{3}}$ and $\mathrm{Sb_{2}Te_{3}}$ with a single {D}irac
  cone on the surface},\ }\href {https://doi.org/10.1038/nphys1270} {\bibfield
  {journal} {\bibinfo  {journal} {Nature Physics}\ }\textbf {\bibinfo {volume}
  {5}},\ \bibinfo {pages} {438–442} (\bibinfo {year} {2009})}\BibitemShut
  {NoStop}%
\bibitem [{\citenamefont {Xia}\ \emph {et~al.}(2009)\citenamefont {Xia},
  \citenamefont {Qian}, \citenamefont {Hsieh}, \citenamefont {Wray},
  \citenamefont {Pal}, \citenamefont {Lin}, \citenamefont {Bansil},
  \citenamefont {Grauer}, \citenamefont {Hor}, \citenamefont {Cava},\ and\
  \citenamefont
  {et~al.}}]{xia_qian_hsieh_wray_pal_lin_bansil_grauer_hor_cava_etal_2009}%
  \BibitemOpen
  \bibfield  {author} {\bibinfo {author} {\bibfnamefont {Y.}~\bibnamefont
  {Xia}}, \bibinfo {author} {\bibfnamefont {D.}~\bibnamefont {Qian}}, \bibinfo
  {author} {\bibfnamefont {D.}~\bibnamefont {Hsieh}}, \bibinfo {author}
  {\bibfnamefont {L.}~\bibnamefont {Wray}}, \bibinfo {author} {\bibfnamefont
  {A.}~\bibnamefont {Pal}}, \bibinfo {author} {\bibfnamefont {H.}~\bibnamefont
  {Lin}}, \bibinfo {author} {\bibfnamefont {A.}~\bibnamefont {Bansil}},
  \bibinfo {author} {\bibfnamefont {D.}~\bibnamefont {Grauer}}, \bibinfo
  {author} {\bibfnamefont {Y.~S.}\ \bibnamefont {Hor}}, \bibinfo {author}
  {\bibfnamefont {R.~J.}\ \bibnamefont {Cava}},\ and\ \bibinfo {author}
  {\bibnamefont {et~al.}},\ }\bibfield  {title} {\bibinfo {title} {Observation
  of a large-gap topological-insulator class with a single {D}irac cone on the
  surface},\ }\href {https://doi.org/10.1038/nphys1274} {\bibfield  {journal}
  {\bibinfo  {journal} {Nature Physics}\ }\textbf {\bibinfo {volume} {5}},\
  \bibinfo {pages} {398–402} (\bibinfo {year} {2009})}\BibitemShut {NoStop}%
\bibitem [{\citenamefont {Zhang}\ and\ \citenamefont
  {Vishwanath}(2010)}]{PhysRevLett.105.206601}%
  \BibitemOpen
  \bibfield  {author} {\bibinfo {author} {\bibfnamefont {Y.}~\bibnamefont
  {Zhang}}\ and\ \bibinfo {author} {\bibfnamefont {A.}~\bibnamefont
  {Vishwanath}},\ }\bibfield  {title} {\bibinfo {title} {Anomalous
  {A}haronov-{B}ohm conductance oscillations from topological insulator surface
  states},\ }\href {https://doi.org/10.1103/PhysRevLett.105.206601} {\bibfield
  {journal} {\bibinfo  {journal} {Phys. Rev. Lett.}\ }\textbf {\bibinfo
  {volume} {105}},\ \bibinfo {pages} {206601} (\bibinfo {year}
  {2010})}\BibitemShut {NoStop}%
\bibitem [{\citenamefont {Egger}\ \emph {et~al.}(2010)\citenamefont {Egger},
  \citenamefont {Zazunov},\ and\ \citenamefont
  {Yeyati}}]{PhysRevLett.105.136403}%
  \BibitemOpen
  \bibfield  {author} {\bibinfo {author} {\bibfnamefont {R.}~\bibnamefont
  {Egger}}, \bibinfo {author} {\bibfnamefont {A.}~\bibnamefont {Zazunov}},\
  and\ \bibinfo {author} {\bibfnamefont {A.~L.}\ \bibnamefont {Yeyati}},\
  }\bibfield  {title} {\bibinfo {title} {Helical luttinger {L}iquid in
  topological insulator nanowires},\ }\href
  {https://doi.org/10.1103/PhysRevLett.105.136403} {\bibfield  {journal}
  {\bibinfo  {journal} {Phys. Rev. Lett.}\ }\textbf {\bibinfo {volume} {105}},\
  \bibinfo {pages} {136403} (\bibinfo {year} {2010})}\BibitemShut {NoStop}%
\bibitem [{\citenamefont {Hong}\ \emph {et~al.}(2014)\citenamefont {Hong},
  \citenamefont {Zhang}, \citenamefont {Cha}, \citenamefont {Qi},\ and\
  \citenamefont {Cui}}]{hong_zhang_cha_qi_cui_2014}%
  \BibitemOpen
  \bibfield  {author} {\bibinfo {author} {\bibfnamefont {S.~S.}\ \bibnamefont
  {Hong}}, \bibinfo {author} {\bibfnamefont {Y.}~\bibnamefont {Zhang}},
  \bibinfo {author} {\bibfnamefont {J.~J.}\ \bibnamefont {Cha}}, \bibinfo
  {author} {\bibfnamefont {X.-L.}\ \bibnamefont {Qi}},\ and\ \bibinfo {author}
  {\bibfnamefont {Y.}~\bibnamefont {Cui}},\ }\bibfield  {title} {\bibinfo
  {title} {One-dimensional helical transport in topological insulator nanowire
  interferometers},\ }\href {https://doi.org/10.1021/nl500822g} {\bibfield
  {journal} {\bibinfo  {journal} {Nano Letters}\ }\textbf {\bibinfo {volume}
  {14}},\ \bibinfo {pages} {2815–2821} (\bibinfo {year} {2014})}\BibitemShut
  {NoStop}%
\bibitem [{\citenamefont {Cho}\ \emph {et~al.}(2015)\citenamefont {Cho},
  \citenamefont {Dellabetta}, \citenamefont {Zhong}, \citenamefont
  {Schneeloch}, \citenamefont {Liu}, \citenamefont {Gu}, \citenamefont
  {Gilbert},\ and\ \citenamefont
  {Mason}}]{cho_dellabetta_zhong_schneeloch_liu_gu_gilbert_mason_2015}%
  \BibitemOpen
  \bibfield  {author} {\bibinfo {author} {\bibfnamefont {S.}~\bibnamefont
  {Cho}}, \bibinfo {author} {\bibfnamefont {B.}~\bibnamefont {Dellabetta}},
  \bibinfo {author} {\bibfnamefont {R.}~\bibnamefont {Zhong}}, \bibinfo
  {author} {\bibfnamefont {J.}~\bibnamefont {Schneeloch}}, \bibinfo {author}
  {\bibfnamefont {T.}~\bibnamefont {Liu}}, \bibinfo {author} {\bibfnamefont
  {G.}~\bibnamefont {Gu}}, \bibinfo {author} {\bibfnamefont {M.~J.}\
  \bibnamefont {Gilbert}},\ and\ \bibinfo {author} {\bibfnamefont
  {N.}~\bibnamefont {Mason}},\ }\bibfield  {title} {\bibinfo {title}
  {Aharonov–{B}ohm oscillations in a quasi-ballistic three-dimensional
  topological insulator nanowire},\ }\bibfield  {journal} {\bibinfo  {journal}
  {Nature Communications}\ }\textbf {\bibinfo {volume} {6}},\ \href
  {https://doi.org/10.1038/ncomms8634} {10.1038/ncomms8634} (\bibinfo {year}
  {2015})\BibitemShut {NoStop}%
\bibitem [{\citenamefont {Wang}\ \emph {et~al.}(2016)\citenamefont {Wang},
  \citenamefont {Li}, \citenamefont {Yu},\ and\ \citenamefont
  {Liao}}]{wang_li_yu_liao_2016}%
  \BibitemOpen
  \bibfield  {author} {\bibinfo {author} {\bibfnamefont {L.-X.}\ \bibnamefont
  {Wang}}, \bibinfo {author} {\bibfnamefont {C.-Z.}\ \bibnamefont {Li}},
  \bibinfo {author} {\bibfnamefont {D.-P.}\ \bibnamefont {Yu}},\ and\ \bibinfo
  {author} {\bibfnamefont {Z.-M.}\ \bibnamefont {Liao}},\ }\bibfield  {title}
  {\bibinfo {title} {Aharonov–{B}ohm oscillations in {D}irac semimetal
  $\mathrm{Cd_{3}As_{2}}$ nanowires},\ }\bibfield  {journal} {\bibinfo
  {journal} {Nature Communications}\ }\textbf {\bibinfo {volume} {7}},\ \href
  {https://doi.org/10.1038/ncomms10769} {10.1038/ncomms10769} (\bibinfo {year}
  {2016})\BibitemShut {NoStop}%
\bibitem [{\citenamefont {Lin}\ \emph {et~al.}(2017)\citenamefont {Lin},
  \citenamefont {Wang}, \citenamefont {Wang}, \citenamefont {Li}, \citenamefont
  {Li}, \citenamefont {Yu},\ and\ \citenamefont {Liao}}]{PhysRevB.95.235436}%
  \BibitemOpen
  \bibfield  {author} {\bibinfo {author} {\bibfnamefont {B.-C.}\ \bibnamefont
  {Lin}}, \bibinfo {author} {\bibfnamefont {S.}~\bibnamefont {Wang}}, \bibinfo
  {author} {\bibfnamefont {L.-X.}\ \bibnamefont {Wang}}, \bibinfo {author}
  {\bibfnamefont {C.-Z.}\ \bibnamefont {Li}}, \bibinfo {author} {\bibfnamefont
  {J.-G.}\ \bibnamefont {Li}}, \bibinfo {author} {\bibfnamefont
  {D.}~\bibnamefont {Yu}},\ and\ \bibinfo {author} {\bibfnamefont {Z.-M.}\
  \bibnamefont {Liao}},\ }\bibfield  {title} {\bibinfo {title} {Gate-tuned
  {A}haronov-{B}ohm interference of surface states in a quasiballistic {D}irac
  semimetal nanowire},\ }\href {https://doi.org/10.1103/PhysRevB.95.235436}
  {\bibfield  {journal} {\bibinfo  {journal} {Phys. Rev. B}\ }\textbf {\bibinfo
  {volume} {95}},\ \bibinfo {pages} {235436} (\bibinfo {year}
  {2017})}\BibitemShut {NoStop}%
\bibitem [{\citenamefont {Ying}\ \emph {et~al.}(2019)\citenamefont {Ying},
  \citenamefont {Yang}, \citenamefont {Lyu}, \citenamefont {Liu}, \citenamefont
  {Ji}, \citenamefont {Fan}, \citenamefont {Yang}, \citenamefont {Jing},
  \citenamefont {Yang}, \citenamefont {Lu},\ and\ \citenamefont
  {Qu}}]{PhysRevB.100.235307}%
  \BibitemOpen
  \bibfield  {author} {\bibinfo {author} {\bibfnamefont {J.}~\bibnamefont
  {Ying}}, \bibinfo {author} {\bibfnamefont {G.}~\bibnamefont {Yang}}, \bibinfo
  {author} {\bibfnamefont {Z.}~\bibnamefont {Lyu}}, \bibinfo {author}
  {\bibfnamefont {G.}~\bibnamefont {Liu}}, \bibinfo {author} {\bibfnamefont
  {Z.}~\bibnamefont {Ji}}, \bibinfo {author} {\bibfnamefont {J.}~\bibnamefont
  {Fan}}, \bibinfo {author} {\bibfnamefont {C.}~\bibnamefont {Yang}}, \bibinfo
  {author} {\bibfnamefont {X.}~\bibnamefont {Jing}}, \bibinfo {author}
  {\bibfnamefont {H.}~\bibnamefont {Yang}}, \bibinfo {author} {\bibfnamefont
  {L.}~\bibnamefont {Lu}},\ and\ \bibinfo {author} {\bibfnamefont
  {F.}~\bibnamefont {Qu}},\ }\bibfield  {title} {\bibinfo {title} {Gate-tunable
  $h/e$-period magnetoresistance oscillations in $\mathrm{Bi_{2}O_{2}Se}$
  nanowires},\ }\href {https://doi.org/10.1103/PhysRevB.100.235307} {\bibfield
  {journal} {\bibinfo  {journal} {Phys. Rev. B}\ }\textbf {\bibinfo {volume}
  {100}},\ \bibinfo {pages} {235307} (\bibinfo {year} {2019})}\BibitemShut
  {NoStop}%
\bibitem [{\citenamefont {Imura}\ and\ \citenamefont
  {Takane}(2013)}]{PhysRevB.87.205409}%
  \BibitemOpen
  \bibfield  {author} {\bibinfo {author} {\bibfnamefont {K.-I.}\ \bibnamefont
  {Imura}}\ and\ \bibinfo {author} {\bibfnamefont {Y.}~\bibnamefont {Takane}},\
  }\bibfield  {title} {\bibinfo {title} {Protection of the surface states in
  topological insulators: Berry phase perspective},\ }\href
  {https://doi.org/10.1103/PhysRevB.87.205409} {\bibfield  {journal} {\bibinfo
  {journal} {Phys. Rev. B}\ }\textbf {\bibinfo {volume} {87}},\ \bibinfo
  {pages} {205409} (\bibinfo {year} {2013})}\BibitemShut {NoStop}%
\bibitem [{\citenamefont {Rosenberg}\ \emph {et~al.}(2010)\citenamefont
  {Rosenberg}, \citenamefont {Guo},\ and\ \citenamefont
  {Franz}}]{PhysRevB.82.041104}%
  \BibitemOpen
  \bibfield  {author} {\bibinfo {author} {\bibfnamefont {G.}~\bibnamefont
  {Rosenberg}}, \bibinfo {author} {\bibfnamefont {H.-M.}\ \bibnamefont {Guo}},\
  and\ \bibinfo {author} {\bibfnamefont {M.}~\bibnamefont {Franz}},\ }\bibfield
   {title} {\bibinfo {title} {Wormhole effect in a strong topological
  insulator},\ }\href {https://doi.org/10.1103/PhysRevB.82.041104} {\bibfield
  {journal} {\bibinfo  {journal} {Phys. Rev. B}\ }\textbf {\bibinfo {volume}
  {82}},\ \bibinfo {pages} {041104} (\bibinfo {year} {2010})}\BibitemShut
  {NoStop}%
\bibitem [{\citenamefont {Keldysh}(1965)}]{SPJ}%
  \BibitemOpen
  \bibfield  {author} {\bibinfo {author} {\bibfnamefont {L.~V.}\ \bibnamefont
  {Keldysh}},\ }\bibfield  {title} {\bibinfo {title} {Diagram technique for
  nonequilibrium processes},\ }\href@noop {} {\bibfield  {journal} {\bibinfo
  {journal} {SOVIET PHYSICS JETP}\ }\textbf {\bibinfo {volume} {20}},\ \bibinfo
  {pages} {1018} (\bibinfo {year} {1965})}\BibitemShut {NoStop}%
\bibitem [{\citenamefont {Zhou}\ \emph {et~al.}(2017)\citenamefont {Zhou},
  \citenamefont {Jiang}, \citenamefont {Xie},\ and\ \citenamefont
  {Sun}}]{PhysRevB.95.245137}%
  \BibitemOpen
  \bibfield  {author} {\bibinfo {author} {\bibfnamefont {Y.-F.}\ \bibnamefont
  {Zhou}}, \bibinfo {author} {\bibfnamefont {H.}~\bibnamefont {Jiang}},
  \bibinfo {author} {\bibfnamefont {X.~C.}\ \bibnamefont {Xie}},\ and\ \bibinfo
  {author} {\bibfnamefont {Q.-F.}\ \bibnamefont {Sun}},\ }\bibfield  {title}
  {\bibinfo {title} {Two-dimensional lattice model for the surface states of
  topological insulators},\ }\href {https://doi.org/10.1103/PhysRevB.95.245137}
  {\bibfield  {journal} {\bibinfo  {journal} {Phys. Rev. B}\ }\textbf {\bibinfo
  {volume} {95}},\ \bibinfo {pages} {245137} (\bibinfo {year}
  {2017})}\BibitemShut {NoStop}%
\bibitem [{\citenamefont {Nielsen}\ and\ \citenamefont
  {Ninomiya}(1981{\natexlab{a}})}]{NIELSEN198120}%
  \BibitemOpen
  \bibfield  {author} {\bibinfo {author} {\bibfnamefont {H.}~\bibnamefont
  {Nielsen}}\ and\ \bibinfo {author} {\bibfnamefont {M.}~\bibnamefont
  {Ninomiya}},\ }\bibfield  {title} {\bibinfo {title} {Absence of neutrinos on
  a lattice: (i). proof by homotopy theory},\ }\href
  {https://doi.org/https://doi.org/10.1016/0550-3213(81)90361-8} {\bibfield
  {journal} {\bibinfo  {journal} {Nuclear Physics B}\ }\textbf {\bibinfo
  {volume} {185}},\ \bibinfo {pages} {20 } (\bibinfo {year}
  {1981}{\natexlab{a}})}\BibitemShut {NoStop}%
\bibitem [{\citenamefont {Nielsen}\ and\ \citenamefont
  {Ninomiya}(1981{\natexlab{b}})}]{NIELSEN1981219}%
  \BibitemOpen
  \bibfield  {author} {\bibinfo {author} {\bibfnamefont {H.}~\bibnamefont
  {Nielsen}}\ and\ \bibinfo {author} {\bibfnamefont {M.}~\bibnamefont
  {Ninomiya}},\ }\bibfield  {title} {\bibinfo {title} {A no-go theorem for
  regularizing chiral fermions},\ }\href
  {https://doi.org/https://doi.org/10.1016/0370-2693(81)91026-1} {\bibfield
  {journal} {\bibinfo  {journal} {Physics Letters B}\ }\textbf {\bibinfo
  {volume} {105}},\ \bibinfo {pages} {219 } (\bibinfo {year}
  {1981}{\natexlab{b}})}\BibitemShut {NoStop}%
\bibitem [{\citenamefont {Kogut}(1983)}]{RevModPhys.55.775}%
  \BibitemOpen
  \bibfield  {author} {\bibinfo {author} {\bibfnamefont {J.~B.}\ \bibnamefont
  {Kogut}},\ }\bibfield  {title} {\bibinfo {title} {The lattice gauge theory
  approach to quantum chromodynamics},\ }\href
  {https://doi.org/10.1103/RevModPhys.55.775} {\bibfield  {journal} {\bibinfo
  {journal} {Rev. Mod. Phys.}\ }\textbf {\bibinfo {volume} {55}},\ \bibinfo
  {pages} {775} (\bibinfo {year} {1983})}\BibitemShut {NoStop}%
\bibitem [{\citenamefont {Marchand}\ and\ \citenamefont
  {Franz}(2012)}]{PhysRevB.86.155146}%
  \BibitemOpen
  \bibfield  {author} {\bibinfo {author} {\bibfnamefont {D.~J.~J.}\
  \bibnamefont {Marchand}}\ and\ \bibinfo {author} {\bibfnamefont
  {M.}~\bibnamefont {Franz}},\ }\bibfield  {title} {\bibinfo {title} {Lattice
  model for the surface states of a topological insulator with applications to
  magnetic and exciton instabilities},\ }\href
  {https://doi.org/10.1103/PhysRevB.86.155146} {\bibfield  {journal} {\bibinfo
  {journal} {Phys. Rev. B}\ }\textbf {\bibinfo {volume} {86}},\ \bibinfo
  {pages} {155146} (\bibinfo {year} {2012})}\BibitemShut {NoStop}%
\bibitem [{\citenamefont {Lu}\ \emph {et~al.}(2011)\citenamefont {Lu},
  \citenamefont {Shi},\ and\ \citenamefont {Shen}}]{PhysRevLett.107.076801}%
  \BibitemOpen
  \bibfield  {author} {\bibinfo {author} {\bibfnamefont {H.-Z.}\ \bibnamefont
  {Lu}}, \bibinfo {author} {\bibfnamefont {J.}~\bibnamefont {Shi}},\ and\
  \bibinfo {author} {\bibfnamefont {S.-Q.}\ \bibnamefont {Shen}},\ }\bibfield
  {title} {\bibinfo {title} {Competition between weak localization and
  antilocalization in topological surface states},\ }\href
  {https://doi.org/10.1103/PhysRevLett.107.076801} {\bibfield  {journal}
  {\bibinfo  {journal} {Phys. Rev. Lett.}\ }\textbf {\bibinfo {volume} {107}},\
  \bibinfo {pages} {076801} (\bibinfo {year} {2011})}\BibitemShut {NoStop}%
\bibitem [{\citenamefont {Blount}(1962)}]{PhysRev.126.1636}%
  \BibitemOpen
  \bibfield  {author} {\bibinfo {author} {\bibfnamefont {E.~I.}\ \bibnamefont
  {Blount}},\ }\bibfield  {title} {\bibinfo {title} {Bloch electrons in a
  magnetic field},\ }\href {https://doi.org/10.1103/PhysRev.126.1636}
  {\bibfield  {journal} {\bibinfo  {journal} {Phys. Rev.}\ }\textbf {\bibinfo
  {volume} {126}},\ \bibinfo {pages} {1636} (\bibinfo {year}
  {1962})}\BibitemShut {NoStop}%
\bibitem [{\citenamefont {Lee}(2009)}]{PhysRevLett.103.196804}%
  \BibitemOpen
  \bibfield  {author} {\bibinfo {author} {\bibfnamefont {D.-H.}\ \bibnamefont
  {Lee}},\ }\bibfield  {title} {\bibinfo {title} {Surface states of topological
  insulators: The {D}irac {F}ermion in curved two-dimensional spaces},\ }\href
  {https://doi.org/10.1103/PhysRevLett.103.196804} {\bibfield  {journal}
  {\bibinfo  {journal} {Phys. Rev. Lett.}\ }\textbf {\bibinfo {volume} {103}},\
  \bibinfo {pages} {196804} (\bibinfo {year} {2009})}\BibitemShut {NoStop}%
\bibitem [{\citenamefont {Takane}\ and\ \citenamefont
  {Imura}(2013)}]{JPSJ.82.074712}%
  \BibitemOpen
  \bibfield  {author} {\bibinfo {author} {\bibfnamefont {Y.}~\bibnamefont
  {Takane}}\ and\ \bibinfo {author} {\bibfnamefont {K.-I.}\ \bibnamefont
  {Imura}},\ }\bibfield  {title} {\bibinfo {title} {Unified description of
  dirac electrons on a curved surface of topological insulators},\ }\href
  {https://doi.org/10.7566/JPSJ.82.074712} {\bibfield  {journal} {\bibinfo
  {journal} {Journal of the Physical Society of Japan}\ }\textbf {\bibinfo
  {volume} {82}},\ \bibinfo {pages} {074712} (\bibinfo {year}
  {2013})}\BibitemShut {NoStop}%
\bibitem [{Note1()}]{Note1}%
  \BibitemOpen
  \bibinfo {note} {Here, we ignore the side surfaces during the calculation.
  Generally, the mobility of the TI material is low. Therefore, in large
  systems, when electrodes are placed on the top and bottom surfaces, the
  electric current flowing through the side surface is negligible.}\BibitemShut
  {Stop}%
\bibitem [{\citenamefont {Datta}(2007)}]{datta_2007}%
  \BibitemOpen
  \bibfield  {author} {\bibinfo {author} {\bibfnamefont {S.}~\bibnamefont
  {Datta}},\ }\href@noop {} {\emph {\bibinfo {title} {Electronic transport in
  mesoscopic systems}}}\ (\bibinfo  {publisher} {Cambridge Univ. Press},\
  \bibinfo {year} {2007})\BibitemShut {NoStop}%
\bibitem [{\citenamefont {Haug}\ and\ \citenamefont
  {Jauho}(2010)}]{haug_jauho_2010}%
  \BibitemOpen
  \bibfield  {author} {\bibinfo {author} {\bibfnamefont {H.}~\bibnamefont
  {Haug}}\ and\ \bibinfo {author} {\bibfnamefont {A.-P.}\ \bibnamefont
  {Jauho}},\ }\href@noop {} {\emph {\bibinfo {title} {Quantum kinetics in
  transport and optics of semiconductors}}}\ (\bibinfo  {publisher}
  {Springer},\ \bibinfo {year} {2010})\BibitemShut {NoStop}%
\bibitem [{\citenamefont {Jiang}\ \emph {et~al.}(2009)\citenamefont {Jiang},
  \citenamefont {Wang}, \citenamefont {Sun},\ and\ \citenamefont
  {Xie}}]{PhysRevB.80.165316}%
  \BibitemOpen
  \bibfield  {author} {\bibinfo {author} {\bibfnamefont {H.}~\bibnamefont
  {Jiang}}, \bibinfo {author} {\bibfnamefont {L.}~\bibnamefont {Wang}},
  \bibinfo {author} {\bibfnamefont {Q.-f.}\ \bibnamefont {Sun}},\ and\ \bibinfo
  {author} {\bibfnamefont {X.~C.}\ \bibnamefont {Xie}},\ }\bibfield  {title}
  {\bibinfo {title} {Numerical study of the topological anderson insulator in
  $\mathrm{HgTe/CdTe}$ quantum wells},\ }\href
  {https://doi.org/10.1103/PhysRevB.80.165316} {\bibfield  {journal} {\bibinfo
  {journal} {Phys. Rev. B}\ }\textbf {\bibinfo {volume} {80}},\ \bibinfo
  {pages} {165316} (\bibinfo {year} {2009})}\BibitemShut {NoStop}%
\bibitem [{\citenamefont {Zhang}\ \emph {et~al.}(2008)\citenamefont {Zhang},
  \citenamefont {Hu}, \citenamefont {Bernevig}, \citenamefont {Wang},
  \citenamefont {Xie},\ and\ \citenamefont {Liu}}]{PhysRevB.78.155413}%
  \BibitemOpen
  \bibfield  {author} {\bibinfo {author} {\bibfnamefont {Y.}~\bibnamefont
  {Zhang}}, \bibinfo {author} {\bibfnamefont {J.-P.}\ \bibnamefont {Hu}},
  \bibinfo {author} {\bibfnamefont {B.~A.}\ \bibnamefont {Bernevig}}, \bibinfo
  {author} {\bibfnamefont {X.~R.}\ \bibnamefont {Wang}}, \bibinfo {author}
  {\bibfnamefont {X.~C.}\ \bibnamefont {Xie}},\ and\ \bibinfo {author}
  {\bibfnamefont {W.~M.}\ \bibnamefont {Liu}},\ }\bibfield  {title} {\bibinfo
  {title} {Quantum blockade and loop currents in graphene with topological
  defects},\ }\href {https://doi.org/10.1103/PhysRevB.78.155413} {\bibfield
  {journal} {\bibinfo  {journal} {Phys. Rev. B}\ }\textbf {\bibinfo {volume}
  {78}},\ \bibinfo {pages} {155413} (\bibinfo {year} {2008})}\BibitemShut
  {NoStop}%
\bibitem [{\citenamefont {Nakanishi}\ and\ \citenamefont
  {Tsukada}(2001)}]{PhysRevLett.87.126801}%
  \BibitemOpen
  \bibfield  {author} {\bibinfo {author} {\bibfnamefont {S.}~\bibnamefont
  {Nakanishi}}\ and\ \bibinfo {author} {\bibfnamefont {M.}~\bibnamefont
  {Tsukada}},\ }\bibfield  {title} {\bibinfo {title} {Quantum loop current in a
  ${C}_{60}$ molecular bridge},\ }\href
  {https://doi.org/10.1103/PhysRevLett.87.126801} {\bibfield  {journal}
  {\bibinfo  {journal} {Phys. Rev. Lett.}\ }\textbf {\bibinfo {volume} {87}},\
  \bibinfo {pages} {126801} (\bibinfo {year} {2001})}\BibitemShut {NoStop}%
\bibitem [{\citenamefont {Aharonov}\ and\ \citenamefont
  {Bohm}(1959)}]{PhysRev.115.485}%
  \BibitemOpen
  \bibfield  {author} {\bibinfo {author} {\bibfnamefont {Y.}~\bibnamefont
  {Aharonov}}\ and\ \bibinfo {author} {\bibfnamefont {D.}~\bibnamefont
  {Bohm}},\ }\bibfield  {title} {\bibinfo {title} {Significance of
  electromagnetic potentials in the quantum theory},\ }\href
  {https://doi.org/10.1103/PhysRev.115.485} {\bibfield  {journal} {\bibinfo
  {journal} {Phys. Rev.}\ }\textbf {\bibinfo {volume} {115}},\ \bibinfo {pages}
  {485} (\bibinfo {year} {1959})}\BibitemShut {NoStop}%
\bibitem [{\citenamefont {Schuster}\ \emph {et~al.}(1997)\citenamefont
  {Schuster}, \citenamefont {Buks}, \citenamefont {Heiblum}, \citenamefont
  {Mahalu}, \citenamefont {Umansky},\ and\ \citenamefont
  {Shtrikman}}]{schuster_buks_heiblum_mahalu_umansky_shtrikman_1997}%
  \BibitemOpen
  \bibfield  {author} {\bibinfo {author} {\bibfnamefont {R.}~\bibnamefont
  {Schuster}}, \bibinfo {author} {\bibfnamefont {E.}~\bibnamefont {Buks}},
  \bibinfo {author} {\bibfnamefont {M.}~\bibnamefont {Heiblum}}, \bibinfo
  {author} {\bibfnamefont {D.}~\bibnamefont {Mahalu}}, \bibinfo {author}
  {\bibfnamefont {V.}~\bibnamefont {Umansky}},\ and\ \bibinfo {author}
  {\bibfnamefont {H.}~\bibnamefont {Shtrikman}},\ }\bibfield  {title} {\bibinfo
  {title} {Phase measurement in a quantum dot via a double-slit interference
  experiment},\ }\href {https://doi.org/10.1038/385417a0} {\bibfield  {journal}
  {\bibinfo  {journal} {Nature}\ }\textbf {\bibinfo {volume} {385}},\ \bibinfo
  {pages} {417–420} (\bibinfo {year} {1997})}\BibitemShut {NoStop}%
\bibitem [{\citenamefont {Klein}(1929)}]{RN82}%
  \BibitemOpen
  \bibfield  {author} {\bibinfo {author} {\bibfnamefont {O.}~\bibnamefont
  {Klein}},\ }\bibfield  {title} {\bibinfo {title} {Die reflexion von
  elektronen an einem potentialsprung nach der relativistischen dynamik von
  {D}irac},\ }\href {https://doi.org/10.1007/BF01339716} {\bibfield  {journal}
  {\bibinfo  {journal} {Zeitschrift für Physik}\ }\textbf {\bibinfo {volume}
  {53}},\ \bibinfo {pages} {157} (\bibinfo {year} {1929})}\BibitemShut
  {NoStop}%
\bibitem [{\citenamefont {Li}\ and\ \citenamefont
  {Shi}(2009)}]{PhysRevB.79.241303}%
  \BibitemOpen
  \bibfield  {author} {\bibinfo {author} {\bibfnamefont {D.}~\bibnamefont
  {Li}}\ and\ \bibinfo {author} {\bibfnamefont {J.}~\bibnamefont {Shi}},\
  }\bibfield  {title} {\bibinfo {title} {{D}orokhov-{M}ello-{P}ereyra-{K}umar
  equation for the edge transport of a quantum spin {H}all insulator},\ }\href
  {https://doi.org/10.1103/PhysRevB.79.241303} {\bibfield  {journal} {\bibinfo
  {journal} {Phys. Rev. B}\ }\textbf {\bibinfo {volume} {79}},\ \bibinfo
  {pages} {241303} (\bibinfo {year} {2009})}\BibitemShut {NoStop}%
\bibitem [{\citenamefont {Kane}\ and\ \citenamefont
  {Mele}(2005)}]{PhysRevLett.95.146802}%
  \BibitemOpen
  \bibfield  {author} {\bibinfo {author} {\bibfnamefont {C.~L.}\ \bibnamefont
  {Kane}}\ and\ \bibinfo {author} {\bibfnamefont {E.~J.}\ \bibnamefont
  {Mele}},\ }\bibfield  {title} {\bibinfo {title} {${Z}_{2}$ topological order
  and the quantum spin {H}all effect},\ }\href
  {https://doi.org/10.1103/PhysRevLett.95.146802} {\bibfield  {journal}
  {\bibinfo  {journal} {Phys. Rev. Lett.}\ }\textbf {\bibinfo {volume} {95}},\
  \bibinfo {pages} {146802} (\bibinfo {year} {2005})}\BibitemShut {NoStop}%
\bibitem [{\citenamefont {Einstein}\ and\ \citenamefont
  {Rosen}(1935)}]{PhysRev.48.73}%
  \BibitemOpen
  \bibfield  {author} {\bibinfo {author} {\bibfnamefont {A.}~\bibnamefont
  {Einstein}}\ and\ \bibinfo {author} {\bibfnamefont {N.}~\bibnamefont
  {Rosen}},\ }\bibfield  {title} {\bibinfo {title} {The particle problem in the
  general theory of relativity},\ }\href
  {https://doi.org/10.1103/PhysRev.48.73} {\bibfield  {journal} {\bibinfo
  {journal} {Phys. Rev.}\ }\textbf {\bibinfo {volume} {48}},\ \bibinfo {pages}
  {73} (\bibinfo {year} {1935})}\BibitemShut {NoStop}%
\bibitem [{\citenamefont {Maier}\ \emph {et~al.}(2017)\citenamefont {Maier},
  \citenamefont {Ziegler}, \citenamefont {Fischer}, \citenamefont {Kozlov},
  \citenamefont {Kvon}, \citenamefont {Mikhailov}, \citenamefont {Dvoretsky},\
  and\ \citenamefont {Weiss}}]{maier}%
  \BibitemOpen
  \bibfield  {author} {\bibinfo {author} {\bibfnamefont {H.}~\bibnamefont
  {Maier}}, \bibinfo {author} {\bibfnamefont {J.}~\bibnamefont {Ziegler}},
  \bibinfo {author} {\bibfnamefont {R.}~\bibnamefont {Fischer}}, \bibinfo
  {author} {\bibfnamefont {D.}~\bibnamefont {Kozlov}}, \bibinfo {author}
  {\bibfnamefont {Z.~D.}\ \bibnamefont {Kvon}}, \bibinfo {author}
  {\bibfnamefont {N.}~\bibnamefont {Mikhailov}}, \bibinfo {author}
  {\bibfnamefont {S.~A.}\ \bibnamefont {Dvoretsky}},\ and\ \bibinfo {author}
  {\bibfnamefont {D.}~\bibnamefont {Weiss}},\ }\bibfield  {title} {\bibinfo
  {title} {Ballistic geometric resistance resonances in a single surface of a
  topological insulator},\ }\bibfield  {journal} {\bibinfo  {journal} {Nature
  Communications}\ }\textbf {\bibinfo {volume} {8}},\ \href
  {https://doi.org/10.1038/s41467-017-01684-0} {10.1038/s41467-017-01684-0}
  (\bibinfo {year} {2017})\BibitemShut {NoStop}%
\bibitem [{\citenamefont {Du}\ \emph {et~al.}(2018)\citenamefont {Du},
  \citenamefont {Wang}, \citenamefont {Scarabelli}, \citenamefont {Pfeiffer},
  \citenamefont {West}, \citenamefont {Fallahi}, \citenamefont {Gardner},
  \citenamefont {Manfra}, \citenamefont {Pellegrini}, \citenamefont {Wind},\
  and\ \citenamefont {et~al.}}]{du_wang}%
  \BibitemOpen
  \bibfield  {author} {\bibinfo {author} {\bibfnamefont {L.}~\bibnamefont
  {Du}}, \bibinfo {author} {\bibfnamefont {S.}~\bibnamefont {Wang}}, \bibinfo
  {author} {\bibfnamefont {D.}~\bibnamefont {Scarabelli}}, \bibinfo {author}
  {\bibfnamefont {L.~N.}\ \bibnamefont {Pfeiffer}}, \bibinfo {author}
  {\bibfnamefont {K.~W.}\ \bibnamefont {West}}, \bibinfo {author}
  {\bibfnamefont {S.}~\bibnamefont {Fallahi}}, \bibinfo {author} {\bibfnamefont
  {G.~C.}\ \bibnamefont {Gardner}}, \bibinfo {author} {\bibfnamefont {M.~J.}\
  \bibnamefont {Manfra}}, \bibinfo {author} {\bibfnamefont {V.}~\bibnamefont
  {Pellegrini}}, \bibinfo {author} {\bibfnamefont {S.~J.}\ \bibnamefont
  {Wind}},\ and\ \bibinfo {author} {\bibnamefont {et~al.}},\ }\bibfield
  {title} {\bibinfo {title} {Emerging many-body effects in semiconductor
  artificial graphene with low disorder},\ }\bibfield  {journal} {\bibinfo
  {journal} {Nature Communications}\ }\textbf {\bibinfo {volume} {9}},\ \href
  {https://doi.org/10.1038/s41467-018-05775-4} {10.1038/s41467-018-05775-4}
  (\bibinfo {year} {2018})\BibitemShut {NoStop}%
\bibitem [{\citenamefont {Yang}\ \emph
  {et~al.}(2019{\natexlab{a}})\citenamefont {Yang}, \citenamefont {Liu},
  \citenamefont {Wang}, \citenamefont {Feng}, \citenamefont {He}, \citenamefont
  {Sun}, \citenamefont {Tang}, \citenamefont {Wu}, \citenamefont {Xiong},
  \citenamefont {Zhang},\ and\ \citenamefont {et~al.}}]{yang_liu}%
  \BibitemOpen
  \bibfield  {author} {\bibinfo {author} {\bibfnamefont {C.}~\bibnamefont
  {Yang}}, \bibinfo {author} {\bibfnamefont {Y.}~\bibnamefont {Liu}}, \bibinfo
  {author} {\bibfnamefont {Y.}~\bibnamefont {Wang}}, \bibinfo {author}
  {\bibfnamefont {L.}~\bibnamefont {Feng}}, \bibinfo {author} {\bibfnamefont
  {Q.}~\bibnamefont {He}}, \bibinfo {author} {\bibfnamefont {J.}~\bibnamefont
  {Sun}}, \bibinfo {author} {\bibfnamefont {Y.}~\bibnamefont {Tang}}, \bibinfo
  {author} {\bibfnamefont {C.}~\bibnamefont {Wu}}, \bibinfo {author}
  {\bibfnamefont {J.}~\bibnamefont {Xiong}}, \bibinfo {author} {\bibfnamefont
  {W.}~\bibnamefont {Zhang}},\ and\ \bibinfo {author} {\bibnamefont {et~al.}},\
  }\bibfield  {title} {\bibinfo {title} {Intermediate bosonic metallic state in
  the superconductor-insulator transition},\ }\href
  {https://doi.org/10.1126/science.aax5798} {\bibfield  {journal} {\bibinfo
  {journal} {Science}\ }\textbf {\bibinfo {volume} {366}},\ \bibinfo {pages}
  {1505–1509} (\bibinfo {year} {2019}{\natexlab{a}})}\BibitemShut {NoStop}%
\bibitem [{\citenamefont {Ozawa}\ \emph {et~al.}(2019)\citenamefont {Ozawa},
  \citenamefont {Price}, \citenamefont {Amo}, \citenamefont {Goldman},
  \citenamefont {Hafezi}, \citenamefont {Lu}, \citenamefont {Rechtsman},
  \citenamefont {Schuster}, \citenamefont {Simon}, \citenamefont {Zilberberg},\
  and\ \citenamefont {Carusotto}}]{RevModPhys.91.015006}%
  \BibitemOpen
  \bibfield  {author} {\bibinfo {author} {\bibfnamefont {T.}~\bibnamefont
  {Ozawa}}, \bibinfo {author} {\bibfnamefont {H.~M.}\ \bibnamefont {Price}},
  \bibinfo {author} {\bibfnamefont {A.}~\bibnamefont {Amo}}, \bibinfo {author}
  {\bibfnamefont {N.}~\bibnamefont {Goldman}}, \bibinfo {author} {\bibfnamefont
  {M.}~\bibnamefont {Hafezi}}, \bibinfo {author} {\bibfnamefont
  {L.}~\bibnamefont {Lu}}, \bibinfo {author} {\bibfnamefont {M.~C.}\
  \bibnamefont {Rechtsman}}, \bibinfo {author} {\bibfnamefont {D.}~\bibnamefont
  {Schuster}}, \bibinfo {author} {\bibfnamefont {J.}~\bibnamefont {Simon}},
  \bibinfo {author} {\bibfnamefont {O.}~\bibnamefont {Zilberberg}},\ and\
  \bibinfo {author} {\bibfnamefont {I.}~\bibnamefont {Carusotto}},\ }\bibfield
  {title} {\bibinfo {title} {Topological photonics},\ }\href
  {https://doi.org/10.1103/RevModPhys.91.015006} {\bibfield  {journal}
  {\bibinfo  {journal} {Rev. Mod. Phys.}\ }\textbf {\bibinfo {volume} {91}},\
  \bibinfo {pages} {015006} (\bibinfo {year} {2019})}\BibitemShut {NoStop}%
\bibitem [{\citenamefont {Lu}\ \emph {et~al.}(2014)\citenamefont {Lu},
  \citenamefont {Joannopoulos},\ and\ \citenamefont
  {Soljačić}}]{lu_joannopoulos_2014}%
  \BibitemOpen
  \bibfield  {author} {\bibinfo {author} {\bibfnamefont {L.}~\bibnamefont
  {Lu}}, \bibinfo {author} {\bibfnamefont {J.~D.}\ \bibnamefont
  {Joannopoulos}},\ and\ \bibinfo {author} {\bibfnamefont {M.}~\bibnamefont
  {Soljačić}},\ }\bibfield  {title} {\bibinfo {title} {Topological
  photonics},\ }\href {https://doi.org/10.1038/nphoton.2014.248} {\bibfield
  {journal} {\bibinfo  {journal} {Nature Photonics}\ }\textbf {\bibinfo
  {volume} {8}},\ \bibinfo {pages} {821–829} (\bibinfo {year}
  {2014})}\BibitemShut {NoStop}%
\bibitem [{\citenamefont {Liu}\ \emph {et~al.}(2019)\citenamefont {Liu},
  \citenamefont {Chen},\ and\ \citenamefont {Xu}}]{liu_chen_xu_2019}%
  \BibitemOpen
  \bibfield  {author} {\bibinfo {author} {\bibfnamefont {Y.}~\bibnamefont
  {Liu}}, \bibinfo {author} {\bibfnamefont {X.}~\bibnamefont {Chen}},\ and\
  \bibinfo {author} {\bibfnamefont {Y.}~\bibnamefont {Xu}},\ }\bibfield
  {title} {\bibinfo {title} {Topological phononics: From fundamental models to
  real materials},\ }\href {https://doi.org/10.1002/adfm.201904784} {\bibfield
  {journal} {\bibinfo  {journal} {Advanced Functional Materials}\ }\textbf
  {\bibinfo {volume} {30}},\ \bibinfo {pages} {1904784} (\bibinfo {year}
  {2019})}\BibitemShut {NoStop}%
\bibitem [{\citenamefont {Chen}\ and\ \citenamefont
  {Wu}(2016)}]{PhysRevApplied.5.054021}%
  \BibitemOpen
  \bibfield  {author} {\bibinfo {author} {\bibfnamefont {Z.-G.}\ \bibnamefont
  {Chen}}\ and\ \bibinfo {author} {\bibfnamefont {Y.}~\bibnamefont {Wu}},\
  }\bibfield  {title} {\bibinfo {title} {Tunable topological phononic
  crystals},\ }\href {https://doi.org/10.1103/PhysRevApplied.5.054021}
  {\bibfield  {journal} {\bibinfo  {journal} {Phys. Rev. Applied}\ }\textbf
  {\bibinfo {volume} {5}},\ \bibinfo {pages} {054021} (\bibinfo {year}
  {2016})}\BibitemShut {NoStop}%
\bibitem [{\citenamefont {Cai}\ \emph {et~al.}(2020)\citenamefont {Cai},
  \citenamefont {Ye}, \citenamefont {Qiu}, \citenamefont {Xiao}, \citenamefont
  {Yu}, \citenamefont {Ke},\ and\ \citenamefont
  {Liu}}]{cai_ye_qiu_xiao_yu_ke_liu_2020}%
  \BibitemOpen
  \bibfield  {author} {\bibinfo {author} {\bibfnamefont {X.}~\bibnamefont
  {Cai}}, \bibinfo {author} {\bibfnamefont {L.}~\bibnamefont {Ye}}, \bibinfo
  {author} {\bibfnamefont {C.}~\bibnamefont {Qiu}}, \bibinfo {author}
  {\bibfnamefont {M.}~\bibnamefont {Xiao}}, \bibinfo {author} {\bibfnamefont
  {R.}~\bibnamefont {Yu}}, \bibinfo {author} {\bibfnamefont {M.}~\bibnamefont
  {Ke}},\ and\ \bibinfo {author} {\bibfnamefont {Z.}~\bibnamefont {Liu}},\
  }\bibfield  {title} {\bibinfo {title} {Symmetry-enforced three-dimensional
  {D}irac phononic crystals},\ }\bibfield  {journal} {\bibinfo  {journal}
  {Light: Science \& Applications}\ }\textbf {\bibinfo {volume} {9}},\ \href
  {https://doi.org/10.1038/s41377-020-0273-4} {10.1038/s41377-020-0273-4}
  (\bibinfo {year} {2020})\BibitemShut {NoStop}%
\bibitem [{\citenamefont {Ningyuan}\ \emph {et~al.}(2015)\citenamefont
  {Ningyuan}, \citenamefont {Owens}, \citenamefont {Sommer}, \citenamefont
  {Schuster},\ and\ \citenamefont {Simon}}]{PhysRevX.5.021031}%
  \BibitemOpen
  \bibfield  {author} {\bibinfo {author} {\bibfnamefont {J.}~\bibnamefont
  {Ningyuan}}, \bibinfo {author} {\bibfnamefont {C.}~\bibnamefont {Owens}},
  \bibinfo {author} {\bibfnamefont {A.}~\bibnamefont {Sommer}}, \bibinfo
  {author} {\bibfnamefont {D.}~\bibnamefont {Schuster}},\ and\ \bibinfo
  {author} {\bibfnamefont {J.}~\bibnamefont {Simon}},\ }\bibfield  {title}
  {\bibinfo {title} {Time- and site-resolved dynamics in a topological
  circuit},\ }\href {https://doi.org/10.1103/PhysRevX.5.021031} {\bibfield
  {journal} {\bibinfo  {journal} {Phys. Rev. X}\ }\textbf {\bibinfo {volume}
  {5}},\ \bibinfo {pages} {021031} (\bibinfo {year} {2015})}\BibitemShut
  {NoStop}%
\bibitem [{\citenamefont {Albert}\ \emph
  {et~al.}(2015{\natexlab{a}})\citenamefont {Albert}, \citenamefont {Glazman},\
  and\ \citenamefont {Jiang}}]{PhysRevLett.114.173902}%
  \BibitemOpen
  \bibfield  {author} {\bibinfo {author} {\bibfnamefont {V.~V.}\ \bibnamefont
  {Albert}}, \bibinfo {author} {\bibfnamefont {L.~I.}\ \bibnamefont
  {Glazman}},\ and\ \bibinfo {author} {\bibfnamefont {L.}~\bibnamefont
  {Jiang}},\ }\bibfield  {title} {\bibinfo {title} {Topological properties of
  linear circuit lattices},\ }\href
  {https://doi.org/10.1103/PhysRevLett.114.173902} {\bibfield  {journal}
  {\bibinfo  {journal} {Phys. Rev. Lett.}\ }\textbf {\bibinfo {volume} {114}},\
  \bibinfo {pages} {173902} (\bibinfo {year} {2015}{\natexlab{a}})}\BibitemShut
  {NoStop}%
\bibitem [{\citenamefont {Yang}\ \emph
  {et~al.}(2019{\natexlab{b}})\citenamefont {Yang}, \citenamefont {Gao},
  \citenamefont {Xue}, \citenamefont {Zhang}, \citenamefont {He}, \citenamefont
  {Yang}, \citenamefont {Singh}, \citenamefont {Chong}, \citenamefont {Zhang},
  \citenamefont {Chen},\ and\ \citenamefont {et~al.}}]{yang_gao_xue_}%
  \BibitemOpen
  \bibfield  {author} {\bibinfo {author} {\bibfnamefont {Y.}~\bibnamefont
  {Yang}}, \bibinfo {author} {\bibfnamefont {Z.}~\bibnamefont {Gao}}, \bibinfo
  {author} {\bibfnamefont {H.}~\bibnamefont {Xue}}, \bibinfo {author}
  {\bibfnamefont {L.}~\bibnamefont {Zhang}}, \bibinfo {author} {\bibfnamefont
  {M.}~\bibnamefont {He}}, \bibinfo {author} {\bibfnamefont {Z.}~\bibnamefont
  {Yang}}, \bibinfo {author} {\bibfnamefont {R.}~\bibnamefont {Singh}},
  \bibinfo {author} {\bibfnamefont {Y.}~\bibnamefont {Chong}}, \bibinfo
  {author} {\bibfnamefont {B.}~\bibnamefont {Zhang}}, \bibinfo {author}
  {\bibfnamefont {H.}~\bibnamefont {Chen}},\ and\ \bibinfo {author}
  {\bibnamefont {et~al.}},\ }\bibfield  {title} {\bibinfo {title} {Realization
  of a three-dimensional photonic topological insulator},\ }\href
  {https://doi.org/10.1038/s41586-018-0829-0} {\bibfield  {journal} {\bibinfo
  {journal} {Nature}\ }\textbf {\bibinfo {volume} {565}},\ \bibinfo {pages}
  {622–626} (\bibinfo {year} {2019}{\natexlab{b}})}\BibitemShut {NoStop}%
\bibitem [{\citenamefont {He}\ \emph {et~al.}(2019)\citenamefont {He},
  \citenamefont {Yu}, \citenamefont {Wang}, \citenamefont {Ge}, \citenamefont
  {Ruan}, \citenamefont {Zhang}, \citenamefont {Lu},\ and\ \citenamefont
  {Chen}}]{PhysRevLett.123.195503}%
  \BibitemOpen
  \bibfield  {author} {\bibinfo {author} {\bibfnamefont {C.}~\bibnamefont
  {He}}, \bibinfo {author} {\bibfnamefont {S.-Y.}\ \bibnamefont {Yu}}, \bibinfo
  {author} {\bibfnamefont {H.}~\bibnamefont {Wang}}, \bibinfo {author}
  {\bibfnamefont {H.}~\bibnamefont {Ge}}, \bibinfo {author} {\bibfnamefont
  {J.}~\bibnamefont {Ruan}}, \bibinfo {author} {\bibfnamefont {H.}~\bibnamefont
  {Zhang}}, \bibinfo {author} {\bibfnamefont {M.-H.}\ \bibnamefont {Lu}},\ and\
  \bibinfo {author} {\bibfnamefont {Y.-F.}\ \bibnamefont {Chen}},\ }\bibfield
  {title} {\bibinfo {title} {Hybrid acoustic topological insulator in three
  dimensions},\ }\href {https://doi.org/10.1103/PhysRevLett.123.195503}
  {\bibfield  {journal} {\bibinfo  {journal} {Phys. Rev. Lett.}\ }\textbf
  {\bibinfo {volume} {123}},\ \bibinfo {pages} {195503} (\bibinfo {year}
  {2019})}\BibitemShut {NoStop}%
\bibitem [{\citenamefont {Wang}\ \emph {et~al.}(2008)\citenamefont {Wang},
  \citenamefont {Chong}, \citenamefont {Joannopoulos},\ and\ \citenamefont
  {Soljačić}}]{RN55}%
  \BibitemOpen
  \bibfield  {author} {\bibinfo {author} {\bibfnamefont {Z.}~\bibnamefont
  {Wang}}, \bibinfo {author} {\bibfnamefont {Y.~D.}\ \bibnamefont {Chong}},
  \bibinfo {author} {\bibfnamefont {J.~D.}\ \bibnamefont {Joannopoulos}},\ and\
  \bibinfo {author} {\bibfnamefont {M.}~\bibnamefont {Soljačić}},\ }\bibfield
   {title} {\bibinfo {title} {Reflection-free one-way edge modes in a
  gyromagnetic photonic crystal},\ }\href
  {https://doi.org/10.1103/PhysRevLett.100.013905} {\bibfield  {journal}
  {\bibinfo  {journal} {Phys Rev Lett}\ }\textbf {\bibinfo {volume} {100}},\
  \bibinfo {pages} {013905} (\bibinfo {year} {2008})}\BibitemShut {NoStop}%
\bibitem [{\citenamefont {Wang}\ \emph {et~al.}(2009)\citenamefont {Wang},
  \citenamefont {Chong}, \citenamefont {Joannopoulos},\ and\ \citenamefont
  {Soljačić}}]{RN32}%
  \BibitemOpen
  \bibfield  {author} {\bibinfo {author} {\bibfnamefont {Z.}~\bibnamefont
  {Wang}}, \bibinfo {author} {\bibfnamefont {Y.}~\bibnamefont {Chong}},
  \bibinfo {author} {\bibfnamefont {J.~D.}\ \bibnamefont {Joannopoulos}},\ and\
  \bibinfo {author} {\bibfnamefont {M.}~\bibnamefont {Soljačić}},\ }\bibfield
   {title} {\bibinfo {title} {Observation of unidirectional
  backscattering-immune topological electromagnetic states},\ }\href
  {https://doi.org/10.1038/nature08293} {\bibfield  {journal} {\bibinfo
  {journal} {Nature}\ }\textbf {\bibinfo {volume} {461}},\ \bibinfo {pages}
  {772} (\bibinfo {year} {2009})}\BibitemShut {NoStop}%
\bibitem [{\citenamefont {Wen}\ \emph {et~al.}(2019)\citenamefont {Wen},
  \citenamefont {Qiu}, \citenamefont {Qi}, \citenamefont {Ye}, \citenamefont
  {Ke}, \citenamefont {Zhang},\ and\ \citenamefont {Liu}}]{RN68}%
  \BibitemOpen
  \bibfield  {author} {\bibinfo {author} {\bibfnamefont {X.}~\bibnamefont
  {Wen}}, \bibinfo {author} {\bibfnamefont {C.}~\bibnamefont {Qiu}}, \bibinfo
  {author} {\bibfnamefont {Y.}~\bibnamefont {Qi}}, \bibinfo {author}
  {\bibfnamefont {L.}~\bibnamefont {Ye}}, \bibinfo {author} {\bibfnamefont
  {M.}~\bibnamefont {Ke}}, \bibinfo {author} {\bibfnamefont {F.}~\bibnamefont
  {Zhang}},\ and\ \bibinfo {author} {\bibfnamefont {Z.}~\bibnamefont {Liu}},\
  }\bibfield  {title} {\bibinfo {title} {Acoustic {L}andau quantization and
  quantum-{H}all-like edge states},\ }\href
  {https://doi.org/10.1038/s41567-019-0446-3} {\bibfield  {journal} {\bibinfo
  {journal} {Nature Physics}\ }\textbf {\bibinfo {volume} {15}},\ \bibinfo
  {pages} {352} (\bibinfo {year} {2019})}\BibitemShut {NoStop}%
\bibitem [{\citenamefont {Albert}\ \emph
  {et~al.}(2015{\natexlab{b}})\citenamefont {Albert}, \citenamefont {Glazman},\
  and\ \citenamefont {Jiang}}]{RN59}%
  \BibitemOpen
  \bibfield  {author} {\bibinfo {author} {\bibfnamefont {V.~V.}\ \bibnamefont
  {Albert}}, \bibinfo {author} {\bibfnamefont {L.~I.}\ \bibnamefont
  {Glazman}},\ and\ \bibinfo {author} {\bibfnamefont {L.}~\bibnamefont
  {Jiang}},\ }\bibfield  {title} {\bibinfo {title} {Topological properties of
  linear circuit lattices},\ }\href
  {https://doi.org/10.1103/PhysRevLett.114.173902} {\bibfield  {journal}
  {\bibinfo  {journal} {Phys Rev Lett}\ }\textbf {\bibinfo {volume} {114}},\
  \bibinfo {pages} {173902} (\bibinfo {year} {2015}{\natexlab{b}})}\BibitemShut
  {NoStop}%
\bibitem [{\citenamefont {Hofmann}\ \emph {et~al.}(2019)\citenamefont
  {Hofmann}, \citenamefont {Helbig}, \citenamefont {Lee}, \citenamefont
  {Greiter},\ and\ \citenamefont {Thomale}}]{RN60}%
  \BibitemOpen
  \bibfield  {author} {\bibinfo {author} {\bibfnamefont {T.}~\bibnamefont
  {Hofmann}}, \bibinfo {author} {\bibfnamefont {T.}~\bibnamefont {Helbig}},
  \bibinfo {author} {\bibfnamefont {C.~H.}\ \bibnamefont {Lee}}, \bibinfo
  {author} {\bibfnamefont {M.}~\bibnamefont {Greiter}},\ and\ \bibinfo {author}
  {\bibfnamefont {R.}~\bibnamefont {Thomale}},\ }\bibfield  {title} {\bibinfo
  {title} {Chiral voltage propagation and calibration in a topolectrical chern
  circuit},\ }\href {https://doi.org/10.1103/PhysRevLett.122.247702} {\bibfield
   {journal} {\bibinfo  {journal} {Phys Rev Lett}\ }\textbf {\bibinfo {volume}
  {122}},\ \bibinfo {pages} {247702} (\bibinfo {year} {2019})}\BibitemShut
  {NoStop}%
\bibitem [{\citenamefont {Brey}\ and\ \citenamefont
  {Fertig}(2014)}]{PhysRevB.89.085305}%
  \BibitemOpen
  \bibfield  {author} {\bibinfo {author} {\bibfnamefont {L.}~\bibnamefont
  {Brey}}\ and\ \bibinfo {author} {\bibfnamefont {H.~A.}\ \bibnamefont
  {Fertig}},\ }\bibfield  {title} {\bibinfo {title} {Electronic states of wires
  and slabs of topological insulators: Quantum {H}all effects and edge
  transport},\ }\href {https://doi.org/10.1103/PhysRevB.89.085305} {\bibfield
  {journal} {\bibinfo  {journal} {Phys. Rev. B}\ }\textbf {\bibinfo {volume}
  {89}},\ \bibinfo {pages} {085305} (\bibinfo {year} {2014})}\BibitemShut
  {NoStop}%
\bibitem [{\citenamefont {Altland~Alexander}(2010)}]{book:826289}%
  \BibitemOpen
  \bibfield  {author} {\bibinfo {author} {\bibfnamefont {S.~B.~D.}\
  \bibnamefont {Altland~Alexander}},\ }\href@noop {} {\emph {\bibinfo {title}
  {Condensed Matter Field Theory, {S}econd Edition}}},\ \bibinfo {edition}
  {2nd}\ ed.\ (\bibinfo  {publisher} {Cambridge University Press},\ \bibinfo
  {year} {2010})\BibitemShut {NoStop}%
\end{thebibliography}%
\end{document}